\documentclass{aa}  
\makeatletter 
\renewcommand*\aa@pageof{, page \thepage{} of \pageref*{LastPage}}
\makeatother

\usepackage{graphicx}
\usepackage[pdfencoding=auto,psdextra]{hyperref}
\hypersetup{
    colorlinks=true,
    linkcolor=blue,
    filecolor=magenta,      
    urlcolor=blue,
    citecolor=blue
}
\usepackage{txfonts}
\usepackage{tablefootnote}
\usepackage{verbatim}
\usepackage{xcolor}
\usepackage{orcidlink}
\usepackage{soul}

\newcommand{\mbh}{$M_{\rm BH}$\xspace}
\newcommand{\ngc}{NGC\,4593\xspace}

\begin{document} 
   \title{Gravitational torques from a lopsided young stellar component sustain high black hole accretion rates in NGC~4593}
   \titlerunning{Gravitational $m=1$ torques in NGC~4593}
    
    \author{N.~Winkel
          \inst{\ref{MPIA}}\orcidlink{0000-0001-9428-6238}
          \and
          K.~Jahnke
          \inst{\ref{MPIA}}\orcidlink{0000-0003-3804-2137}
          \and 
          J.\,A.~Fernández-Ontiveros
          \inst{\ref{CEFCA}}\orcidlink{0000-0001-9490-899X}\:
          \and
          T.~A.~Davis\inst{\ref{Cardiff}}\orcidlink{0000-0003-4932-9379}\:
          \and
          F.~Combes\inst{\ref{Paris}}\orcidlink{0000-0003-2658-7893}\:
          \and
          M.~Gaspari
          \inst{\ref{UNIMORE}}\orcidlink{0000-0003-2754-9258}\:
          \and
          J.~Neumann\inst{\ref{MPIA}}\orcidlink{0000-0002-3289-8914}\:
          \and
          M.~Singha\inst{\ref{NASA_Goddard}, \ref{Catholic_Uni}}\orcidlink{0000-0001-5687-1516}\:
          \and
          J.~S.~Elford\inst{\ref{UNAM}}\orcidlink{0000-0002-6139-2226}\:
          \and
          V.~N.~Bennert\inst{\ref{CalPoly}}\orcidlink{0000-0003-2064-0518}\:
          M.~A.~Malkan\inst{\ref{UCLA}}
          \orcidlink{0000-0001-6919-1237}\:
          }

   \institute{
      Max-Planck-Institut f\"ur Astronomie, K\"onigstuhl 17, D-69117 Heidelberg, Germany, \email{winkel@mpia.de} \label{MPIA}
      \and
      Centro de Estudios de Física del Cosmos de Aragón (CEFCA), Plaza San Juan 1, 44001 Teruel, Spain \label{CEFCA}
      \and 
      Cardiff Hub for Astrophysics Research \&\ Technology, School of Physics \& Astronomy, Cardiff University, CF24 3AA, UK \label{Cardiff}
      \and
      LUX, Observatoire de Paris, PSL Univ., Coll\`ege de France, CNRS, Sorbonne Univ., Paris, France \label{Paris}
      \and
      Department of Physics, Informatics \& Mathematics, University of Modena \& Reggio Emilia, 41125 MO, Italy \label{UNIMORE}
      \and
      Astrophysics Science Division, NASA, Goddard Space Flight Center, Greenbelt, MD 20771, USA \label{NASA_Goddard}
      \and
      Department of Physics, The Catholic University of America, Washington, DC 20064, USA \label{Catholic_Uni}
      \and
      Instituto de Estudios Astrof\'{i}sicos, Facultad de Ingenier\'{i}a y Ciencias, Universidad Diego Portales, Av. Ej\'{e}rcito Libertador 441, Santiago, Chile \label{UNAM}
      \and
      Physics Department, California Polytechnic State University, San Luis Obispo, CA 93407, USA \label{CalPoly}
      \and
      Department of Physics and Astronomy, UCLA, Los Angeles, CA 90095, USA       \label{UCLA}
      }


 
  \abstract
   {Supermassive black holes (SMBHs) grow primarily through gas accretion, observed as active galactic nuclei (AGNs). 
   While mergers can drive luminous AGN episodes, secular processes may fuel a substantial portion of cosmic black hole growth. Whether these mechanisms can sustain high black hole accretion rates remains uncertain.
   }
   {The aim of this study is to identify the secular mechanism driving high SMBH accretion rates, by targeting a galaxy with a moderately massive SMBH, high central gas densities, accretion rates of a few percent of the Eddington limit, and gas kinematics resolved close to black hole-dominated scales.}
   {A blind search led to the identification of \ngc, which is representative of the AGN population driving BH mass density growth since $z=1$. Combining HST imaging, VLT/MUSE spectroscopy, and ALMA imaging, we resolve molecular and ionised gas kinematics close to the sphere of influence of the SMBH.}
   {A prominent single-arm ("$m=1$”) molecular gas spiral with ${\rm log}\,M_{\rm mol}/{\rm M}_\odot=8.1\pm0.3$ extends from 1.3\,kpc down to the SMBH's sphere of influence ($1.7^{+0.5}_{-0.2}\,{\rm pc}$).
   Star formation in the spiral is inefficient (${\rm SFR} = 4.9 \times 10^{-2} \,{\rm M}_\odot {\rm /yr}$, $\langle t_{\rm dep}\rangle=3.9 \pm 0.6\,\rm{Gyr}$), whereas inflow rate exceeds the SFR by two orders of magnitude and is sufficient to sustain the current SMBH accretion rate for $\geq$35\,Myr, enabling $\sim$10\% SMBH growth.
   A young, lopsided stellar component (${\rm log}\,M_\star/{\rm M}_\odot=7.5-9.3$) exerts torques on the molecular gas, likely driving the gas inflow. This young stellar component may serve as both a cause and a product of sustained gas funnelling towards the SMBH.}
   {These findings provide evidence for the sustained secular $m=1$ feeding mode at high SMBH accretion rates, linking kpc-scale gas dynamics to the black hole’s sphere of influence. This mechanism, consistent with simulation predictions, may represent a key contributor to SMBH growth in luminous AGNs since cosmic noon.}
   
   \keywords{galaxies: kinematics and dynamics - galaxies: ISM - galaxies: active - galaxies: Seyfert - galaxies: nuclei - quasars: supermassive black holes}

   \maketitle
%

\section{Introduction}
\label{sec:Intro}
Supermassive black holes (SMBHs) are located in the hearts of massive galaxies. Their enormous mass growth over cosmic time is governed by accretion of material as manifested in bright active galactic nuclei (AGNs). To sustain the AGN phase, gas must be transported from tens of kiloparsecs to well within the very central parsec, a process that involves multiple scales, gas phases, and physical mechanisms (\citealt{Storchi-Bergmann:2019,Gaspari:2020}, for reviews).
Major galaxy mergers provide an efficient channel to transport gas. They are known to trigger luminous quasar phases, i.e.\ AGNs with the highest specific accretion rates.
However, major galaxy mergers are often not required conditions for black hole fuelling: 
Since cosmic noon, the bulk of black hole mass density growth does not seem to have been associated with galaxy merging (see e.g. evidence at $z=1$ in COSMOS \citealt{Cisternas:2011} and $z=2$, in a HST snapshot programme for SDSS quasars \citealt{Mechtley:2016}). 
In addition, half of the AGNs that dominate black hole growth since $z=1$ \citep{Cisternas:2011}, and likely even $z=2$ \citep{Kocevski:2012, Cisternas:2015}, have disc-dominated host galaxies. This implies that these galaxies have had no recent strong interactions with massive companions.
As a consequence, secular or instability processes must also enable gas inflow from kiloparsec to sub-parsec scales.

Some of the detailed galaxy-intrinsic gas transport mechanisms have been resolved in nearby low-luminosity (low-$L$) AGNs, with $L_{\rm bol} \sim 10^{41} - 10^{43}\,{\rm erg\,s}^{-1}$, (corresponding to $10^{-2} - 10^{-5}\, {\rm M}_\odot \,{\rm yr}^{-1}$ assuming thin disc accretion with a standard radiative efficiency $\epsilon \sim 0.1$).
On galaxy scales, bars can efficiently transfer angular momentum within the galaxy disc, allowing gas to radially migrate \citep{Regan&Teuben:2004, Kim:2012, Combes:2014, Sormani:2015}. Within the bar's inner Lindblad resonance (ILR), which is often at the inner edge of a gap that forms around the ILR (typically several 100\,pc from the centre), inflowing gas frequently stalls and forms a resonant ring \citep{Buta&Combes:1996, Regan:1999, Combes:2019, Sormani:2024}. Within the ILR, bars-within-bars contribute gravitational torques \citep[e.g.][]{Schlosman:1989, Maciejewski:2004, Emsellem:2015}, while dynamical friction between colliding gas clouds can provide pressure torques \citep[chaotic cold accretion;][]{Bournaud:2011, Gaspari:2015}.
However, the processes that fuel such low-$L$ AGNs are generally inefficient. The majority of cosmic SMBH growth likely occurred in high-accretion rate AGNs, which accrete at several percent of the Eddington limit. However, the associated gas transport mechanisms within the central kiloparsec that sustain such high black hole accretion rates (BHARs) remain unsettled observationally.

Detailed simulations have investigated torques at small scales. For instance, \citet[][HQ10 hereafter]{Hopkins&Quataert:2010}, simulated the propagation of sustained gas-density instabilities in the central few hundred parsecs of gas-rich galaxies. Their zoom-in simulations predicted that, within the ILR of the galaxy-scale bar, nested gaseous structures exert torques on gas in the inner $\sim$10\,pc through eccentric discs or single-arm (`$m=1$') spirals.
As gas flows inwards, stars rapidly form out of the disc, imprinting the $m=1$ mode in both the stellar and gaseous components. The lopsided mass distributions then precess around the black hole relative to each other, driving continued gas inflow down to sub-parsec scales. Although up to 99\% of the gas may be converted into stars along the way, a substantial amount still reaches the accretion disc, enough to power luminous AGNs.
Hydrodynamic simulations by \cite{Emsellem:2015} also report similar $m=1$ gas features on $\sim$10\,pc scales, where star formation accompanies steady gas accretion. 
By including a sub-grid accretion model based on torque accretion in cosmological hydrodynamic simulations, the rate at which gravitational torques feed the central black hole has been shown to shape the black hole-host galaxy scaling relations \citep{Angles-Alcazar:2017}. Therefore, understanding these gravitational torques in galaxy centres is crucial, as they may be the primary driver of the co-evolution of galaxies and their central black holes.

To detect the characteristic signatures that govern high BHARs, we conducted a blind search for objects that are expected to show observational signatures of the $m=1$ mode.
We imposed (a) \mbh $\sim 10^7$--$10^8\,{\rm M}_\odot$, (b) an Eddington ratio $\lambda_{\rm Edd}$ of several percent, and (c) central molecular gas overdensities that could be resolved down to the black hole's gravitational sphere of influence (SOI).
We identified \ngc (aka Mrk~1330) as one of the rare nearby luminous type-1 AGNs with a high specific accretion rate of $\lambda_{\rm Edd} = 0.06$ \citep{Husemann:2022}. 
The spiral host galaxy of \ngc has a stellar mass of $M_\star = 10^{10.9}\,{\rm M}_\odot$ and a neutral hydrogen mass of $M_{\rm HI} = 10^{9.31}\,{\rm M}_\odot$ \citep{Diaz-Garcia:2021}. \cite{Gadotti:2008} classified \ngc's host as SBb(rs), with a 9.1\,kpc-scale bar \citep{Treuthardt:2012} that contributes 16\% to its total $R$-band luminosity.
A single-epoch black hole mass estimate places \ngc slightly below the lower limit of criterion (a) \citep{Husemann:2022}, as confirmed through reverberation mapping of the broad-line region (BLR) lags and accretion disc reverberation mapping \citep{McHardy:2018}.
Dynamical modelling of velocity-resolved BLR lags constrained \ngc's BLR geometry to a thick disc with an opening angle of $\theta_0 = 43^{+19}_{-22}\,^{\circ}$, inclined by $\theta_i = 32^{+10}_{-19}\,^{\circ}$, which is similar to the host galaxy inclination of $47\,^{\circ}$ \citep{Kianfar:2024}. The associated direct black hole mass measurement is $M_{\rm BH} = 4.47^{+3.85}_{-1.30} \times 10^6\,{\rm M}_\odot$ \citep{Williams:2018}.
\ngc lies slightly below the \mbh-host galaxy scaling relations (e.g. the \mbh-$\sigma_\star$ relation; \citealt{Winkel:2025}), indicating that it hosts an undermassive SMBH.
A single low-mass companion -- likely only weakly interacting, if at all (light ratio 1:7.7, distance $\sim$2 disc radii \citealt{Kollatschny:1985}) -- implies that the sub-kiloparsec gas dynamics are dominated by processes originating from inside the galaxy rather than from an external perturber.
This makes \ngc representative of the luminous AGN population that dominated the growth of the cosmic SMBH mass density since $z=1$ \citep{Merloni:2004,Schulze:2015}.
Due to its proximity, the nuclear molecular gas overdensities in \ngc's centre can be resolved from kiloparsec galaxy scales down to the parsec-scale SOI, making it an ideal target to resolve the processes contributing to the growth of its central black hole.

The goal of this work is to constrain the galaxy-driven accretion processes in \ngc that contribute to providing its AGN with gas from the central 2\,kpc. 
To trace the processes near the galaxy nucleus, we combined multi-wavelength data from Hubble Space Telescope (HST), the Very Large Telescope (VLT), and the Atacama Large Millimeter Array (ALMA) to obtain a multi-chromatic view of the AGN feeding mechanism, from galaxy scales down to the central few parsecs.
Throughout this paper, we assume a flat Lambda cold dark matter cosmology with $H_0 = 70\,\rm{km\,s}^{-1}\,\rm{Mpc}^{-1}$, $\Omega_M = 0.3$, and $\Omega_\Lambda=0.7$. 
Based on \ngc's spectroscopic redshift of $z=0.0085 \pm 0.0003$, as measured in Sect.~\ref{subsec:Analysis_Spectral_Synthesis}, we adopted a luminosity distance of 36.6\,Mpc.

\section{Data}
\label{sec:Data}

\begin{figure*}
    \centering
    \resizebox{.95\hsize}{!}{\includegraphics{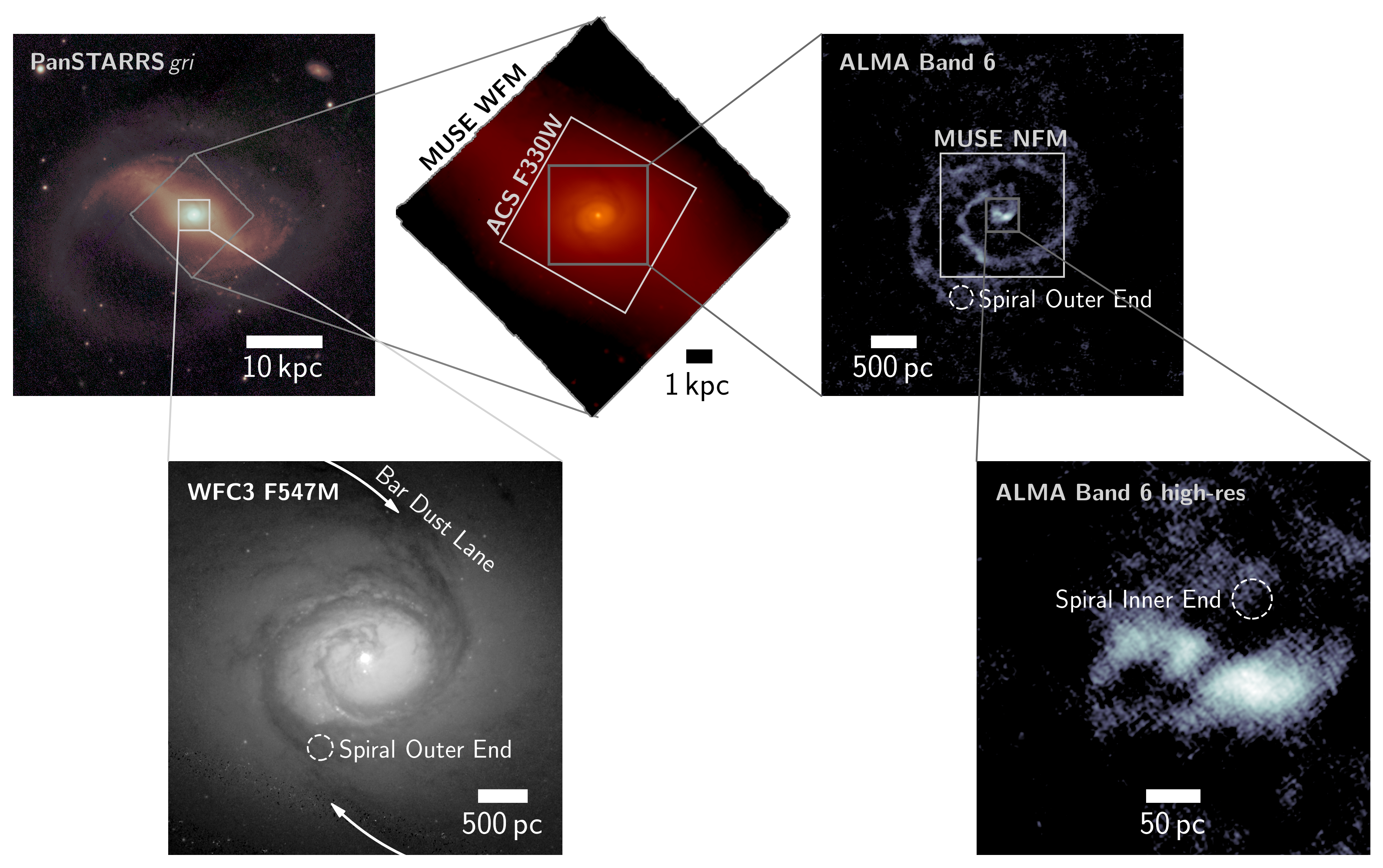}}
         \caption{
         Overview of \ngc's host galaxy on different spatial scales.
         Optical imaging from PanSTARRS \emph{gri} colour (top left) shows the spiral host of \ngc with a the 9.1\,kpc dominant bar. A zoom-in of the HST/WFC3 F547M image shows the prominent dust lanes connecting the dust lanes of the bar with a nuclear single-arm spiral. 
         The MUSE WFM FOV (top centre) covers 1$^\prime \times$1$^\prime$ of the dominant bar. UV imaging from HST/ACS HRC covers the inner 29$\arcsec\times$26$\arcsec$.
          The right panels show the moment 0 map of the ALMA CO(2-1) data combining data taken in two different configurations and covering the innermost 26$\arcsec$ at a resolution of up to 40\,mas.
         }
 \label{fig:Data_Overview}
\end{figure*}

\subsection{ALMA imaging}
\label{subsec:Data_ALMA}

The ALMA CO(2--1) imaging used in this work comes from two different observational programs. Prog.ID 2017.1.00236.S (PI: Malkan) observed \ngc in a setup with lower angular resolution. Using CASA v5.4.0-70, we continuum-subtracted and reconstructed the cube at 10\,km\,s${^{-1}}$ resolution, yielding a major axis of the synthesized beam $b_{\rm maj} = 0\farcs23$ (39\,pc). In addition, our team has observed \ngc under Prop.ID 2018.1.00978.S (PI: Jahnke) in a configuration with higher spatial resolution $b_{\rm maj} = 40\,{\rm mas}$ (6.7\,pc, $\sim 4 \times R_{\rm SOI}$). We created a data cube where we jointly image the two data sets using a Briggs robust parameter of 0.5, which provides an optimal balance between spatial resolution and sensitivity. The cube was constructed with a frequency resolution equivalent to 30\,km\,s${^{-1}}$. We used this combined cube, since it covers the molecular gas structure embedded within the bar at the highest spatial resolution close to the AGN.  The combined cube has an rms noise of $\sim 8 \times 10^{-4}$\,Jy/beam per $30\,{\rm km\,s}^{-1}$ channel, corresponding to $\Sigma_{\rm mol,gas} \approx 0.041 \,{\rm M}_\odot \,{\rm pc}^{-2}$ per $30\,{\rm km\,s}^{-1}$ interval. The CO-to-H$_2$ conversion factor depends on metallicity, density, temperature, and opacity \citep{Bolatto:2013}. 
In galaxy centres, it is often reported to be 4-15$\times$ lower than the Milky Way's canonical value, $\alpha_{\rm CO}^{\rm MW} = 4.35\,{\rm M}_\odot \, {\rm pc}^{-2}\,({\rm K\,km/s})^{-1}$ \citep{Hitschfeld:2008, Sliwa:2013, Zhang:2014, Israel:2020, Teng:2023}. Unless stated differently, we assume a constant $\alpha_{\rm CO} = 1/10 \, \alpha_{\rm CO}^{\rm MW}$ for \ngc's nuclear single-arm spiral, yielding a conservative estimate of the molecular gas masses.

\subsection{Hubble Space Telescope imaging}
\label{subsec:Data_HST_ACS}

We collected archival high-resolution UV imaging of \ngc obtained with HST.
\ngc was observed with the High Resolution Camera (HRC) of the Advanced Camera for Surveys (ACS), using the F330W filter under Prop.ID 9379 on 2003-03-25 (PI: Schmitt). The program includes snapshots of the nuclear regions of AGN hosts, which were designed to detect faint star forming regions. The \ngc data set consists of a long exposure (1140\,s) and a short exposure (60\,s) taken at the same dither location. We retrieved the data from the HST archive, processed through the standard calibration pipeline. 
We corrected the few saturated pixels in the long exposure, located at the centre of the bright point-spread function (PSF), by replacing them with the scaled pixels of the short exposure. Then, we used the \texttt{L.A.Cosmic} algorithm \citep{vanDokkum:2001} routine to remove cosmic rays. The final image has a pixel scale of 0\farcs025 per pixel, covering \ngc's innermost 28\farcs4\,$\times$\,29\farcs1.

\subsection{Optical 3D spectroscopy}
\label{subsec:Data_MUSE}

\ngc was observed with MUSE in both the wide field mode (WFM) and narrow field mode (NFM).
For the WFM observations obtained under Prog.ID 099.B-0242(B) (PI: Carollo), we collected the reduced and calibrated data from the ESO Science archive.
The archival data cube has a field of view (FOV) of 64\arcsec\,$\times$\,65\arcsec\ with a sampling of 0\farcs2 per pixel and a seeing-limited resolution of 1\farcs03. Its wavelength coverage extends from $4750\,\textrm{\r{A}}$ to $9300\,\textrm{\r{A}}$ with a spectral resolution of $\sim$\,2.5\,\r{A}, corresponding to a resolving power that increases from $\mathcal{R}=1750$ (5000\,\AA) to $\mathcal{R}=3500$ (9300\,\AA) \citep{Bacon:2017,Guerou:2017}.

In addition, we observed the central region of \ngc with the adaptive-optics-assisted NFM of the MUSE Integral Field Unit under the program 0103.B-0908(A) (PI: Jahnke). The data were acquired in service mode on the night of 28 Apr 2019 night, with 8 individual exposures of 600\,s each, dithered by 0\farcs5 to minimise the imprint of cosmic rays and flat-fielding artefacts. For the reduction of the data, we have used the MUSE pipeline v2.8.3 together with the graphical user interface \texttt{ESO Reflex} v2.11.0 to execute the \texttt{EsoRex} Common Pipeline Library reduction recipes. We corrected the differential atmospheric refraction as described in \citet{Winkel:2022}, to achieve the highest spatial resolution after combining the individual exposures. The final data cube consists of 374\,$\times$\,367 spaxels, corresponding to a FOV of 9\farcs35\,$\times$\, 9\farcs16 with 137\,258 spectra. Across the 5577\,\r{A}--9350\,\r{A} range, the telluric emission lines yield a nearly constant resolution of \mbox{${\rm FWHM}=2.54 \pm 0.10\,\textrm{\r{A}}$}. This corresponds to 160.4\,km\,s$^{-1}$ ($\mathcal{R}\approx 2190$) and 81.5\,km\,s$^{-1}$ ($\mathcal{R}\approx 3680$), consistent with the MUSE WFM spectral resolution.

\subsection*{Deblending AGN and host emission}
\label{subsubsec:Data_MUSE_Deblending}

For an accurate extraction of the emission line parameters in type-1 AGNs, it is essential to clean the extended host galaxy emission from the point-like AGN emission. This is particularly important close to the galaxy nucleus, where the AGN outshines the host galaxy by orders of magnitude. To achieve a deblending in both spatial and wavelength dimensions, we followed the approach described in \citet{Winkel:2022} where empirical PSF images are measured from the broad emission lines using \texttt{QDeblend}$^{\rm 3D}$ \citep[][]{Husemann:2013}. We generated a hybrid PSF model, consisting of an empirical core, and an analytic model for the outer regions where the signal-to-noise ratio (S/N) is low. We then interpolated in wavelength to construct a wavelength-dependent 3D PSF cube. As a last step, we iteratively subtracted the PSF cube from the original cube. This leaves us with two deblended cubes; the AGN cube contains the point-like emission from the AGN including the power-law continuum and the broad line emission, whereas the host galaxy cube contains the spatially resolved emission, i.e.\ the host galaxy emission.

\section{Analysis and results}
\label{sec:Analysis_and_Results}

\subsection{The single-arm spiral}
\label{subsec:the_single_arm_spiral}
\ngc hosts a striking nuclear single-arm spiral that is clearly traced by dust absorption features in optical continuum images (Fig.~\ref{fig:Data_Overview}, see also \citealt{Kianfar:2024}). This dust is co-spatial with an abundance of molecular gas, as detected in CO(2-1) emission, where the morphology of the single-arm spiral is even more evident. From the CO(2-1) luminosity, we estimate a total molecular gas mass of ${\rm log}\,M_{\rm mol}/{\rm M}_\odot = (0.8 - 4) \times 10^8 \, {\rm M}_\odot$, depending on the CO conversion factor ($\alpha_{\rm CO} = 1/10 \, \alpha_{\rm CO}^{\rm MW}$ vs. $1/2 \,\alpha_{\rm CO}^{\rm MW}$). For the remainder of this analysis, we will adopt the lower boundary. The spiral structure connects the region where the inflowing gas stalls in the ring, near the location where the x1 and x2 orbits cross, at a radius of 1.9\,kpc, to the AGN. From 1.9\,kpc inward, two prominent dust lanes lead towards the centre. Notably, inside a radius of 1.3\,kpc, the spiral transitions to a clearly single-armed structure. This single-arm spiral extends inward over 2.5 windings (see Fig.~\ref{fig:KinMS_modelling}), reaching close to the black hole’s SOI (1.7\,pc, see Sect.~\ref{subsec:Analysis_Spectral_Synthesis}). Recently, \citet{Kianfar:2024} detected non-axisymmetric motions in \ngc's single-arm nuclear gas spiral (see their Fig.~9). They proposed that 5\% of the molecular gas in the spiral is outflowing, but only at a specific location $\sim$220\,pc from the nucleus, qualitatively matching the $<340\,{\rm pc}$ ionized wind east from the centre \citep{Mulumba:2024}. However, molecular gas velocity differences are small (<50\,km\,s$^{-1}$) and could also be attributed to inflows or simply non-circular orbits. With the superior resolution of the ALMA Band 6 data set, we here further diagnose and classify the gas transport processes along this single-arm spiral.

\subsection{Spectral synthesis modelling}
\label{subsec:Analysis_Spectral_Synthesis}
To extract the host galaxy stellar kinematic and emission line parameters, we used the publicly available spectral synthesis modelling code \texttt{PyParadise}\footnote{\url{https://git.io/pyparadise}} \citep{Husemann:2016a, Husemann:2022}. We followed the procedure outlined in \citet{Winkel:2022}. The PSF subtraction described in Sect.~\ref{subsubsec:Data_MUSE_Deblending} leaves strong non-physical continuum variations close to the AGN, artefacts that \texttt{PyParadise} can reliably correct. The \texttt{PyParadise} fitting methodology and its relevance specific to the WFM and NFM data sets are described in \citet{Husemann:2022} and \citet{Winkel:2022}. For \ngc, in brief, we first used the adaptive Voronoi tessellation and binning routine of \citet{Cappellari:2003} to achieve a minimum S/N of 20 in the stellar continuum between 5080--5260\,\r{A}. Next, we modelled the binned stellar continuum spectra using the updated CB09 version of the evolutionary synthesis model spectra from \citet{Bruzual:2003}. To model the emission lines, we tied the stellar kinematics to the measurements obtained in the previous step. This approach ensured improved reliability of the emission line measurements, especially where lines overlap with absorption features.

To model the emission lines from the residual spectrum, we set up \texttt{PyParadise} to use a set of Gaussian models. For the doublet emission lines [\ion{N}{ii}]$\lambda\lambda$6548,83 and [\ion{O}{iii}]$\lambda\lambda$4959,5007, we fixed the flux ratio to the theoretical prediction of 2.96 \citep{Storey:2000,Dimitrijevic:2007}. Furthermore, we coupled the emission lines in radial velocity and velocity dispersion in order to increase the robustness of the flux measurements. The integrated flux remained consistent within uncertainties, regardless of whether emission lines with the same ionisation potentials were kinematically tied. Close to the AGN, however, coupling their velocities becomes crucial for disentangling emission lines from artificial features caused by PSF subtraction. Since the MUSE WFM and NFM observations were conducted under different atmospheric conditions, we analyse the two data sets independently. After analysing the two data sets independently, we combine them in the image plane by degrading the NFM spatial resolution to match that of the seeing-limited WFM data.

To estimate the systemic redshift, which is reported in Sect.~\ref{sec:Intro} and used throughout this work, we extracted a spectrum from a 3\arcsec (520\,pc) aperture of the MUSE WFM data cube. We then fitted the \ion{Ca}{ii}$\lambda\lambda\lambda8498,\,8542,\,8662$ (Ca\,II triplet) stellar absorption lines, yielding a systemic velocity of $2548 \pm 90\,{\rm km\,s^{-1}}$ ($z=0.0085 \pm 0.0003$).

For an estimation of the SOI size of \ngc's SMBH we used $R_{\rm SOI} = G M_{\rm BH} /\sigma_\star^2$. This requires knowledge of the host stellar velocity dispersion at the location of the black hole, which we measure from an aperture spectrum at the AGN location. This spectrum was obtained by integrating the AGN-subtracted data cube over the central $0\farcs2$ (34.4\,pc) of the MUSE NFM data cube, the smallest aperture with an S/N > 20 for the Ca\,II triplet. By fitting the Ca\,II triplet, we derive a stellar velocity dispersion of $\sigma_\star = 105 \pm 12 \,{\rm km\,s^{-1}} $. Using the dynamically measured black hole mass of $M_{\rm BH} = 4.47^{+3.85}_{-1.30} \times 10^6 \,{\rm M}_{\odot}$ \citep{Williams:2018}, we calculate a SOI radius of $R_{\rm SOI} = 1.7 ^{+0.5}_{-0.2}\,{\rm pc}$, slightly below what can be resolved with the high-resolution ALMA data set.

\subsection{Star formation rates}
\label{subsec:star_formation_rates}
To estimate star formation along the single-arm spiral structure in \ngc, we used the emission line maps retrieved from the spectral synthesis analysis described in Sect.~\ref{subsec:Analysis_Spectral_Synthesis}. The H$\alpha$ fluxes were corrected for extinction assuming case B recombination, using an intrinsic Balmer decrement of H$\alpha$/H$\beta$ = 2.86, an electron temperature of $T_e = 10^4 \,{\rm K}$, and density $n_e = 100\,{\rm cm}^{-3}$ with
\begin{equation}
    {\rm H} \alpha_{\rm corr} = {\rm H} \alpha_{\rm obs} \left( \frac{{\rm H} \alpha/{\rm H} \beta}{2.86} \right)^{ \frac{\kappa_\alpha}{\kappa_\beta - \kappa_\alpha}}\,,
\end{equation} 
where $\kappa_\alpha = 2.52$, $\kappa_\beta = 3.66$ \citep{ODonnell:1994}, and Milky Way $R_V = 3.1$. To isolate star formation from AGN ionisation, we employed the Baldwin, Phillips, and Terlevich (BPT) diagram and modelled the mixing sequence using \texttt{Rainbow} \citep{Smirnova-Pinchukova:2022}\footnote{\url{https://gitlab.com/SPIrina/rainbow}} which estimates the star-forming fraction $f_{\rm SF}$ via a likelihood-based comparison to template AGN and SF emission line ratios. \texttt{Rainbow} maximises a likelihood function over the line ratio parameter space,  returning posterior probability distributions for $f_{\rm SF}$ and its associated uncertainty. For \ngc, we identified two reference regions: a central AGN-dominated core within $<$1\arcsec, and a spiral-arm segment with line ratios consistent with pure star formation. These anchor points constrain the model and allow for robust interpolation across mixed-excitation zones.
Across the spaxels spanning the range between these two reference regions, the star-forming ionisation fraction $f_{\rm SF}$ and its uncertainty are considered free parameters, and are determined from the probability distribution.
\begin{figure*}
 \centering
 \includegraphics{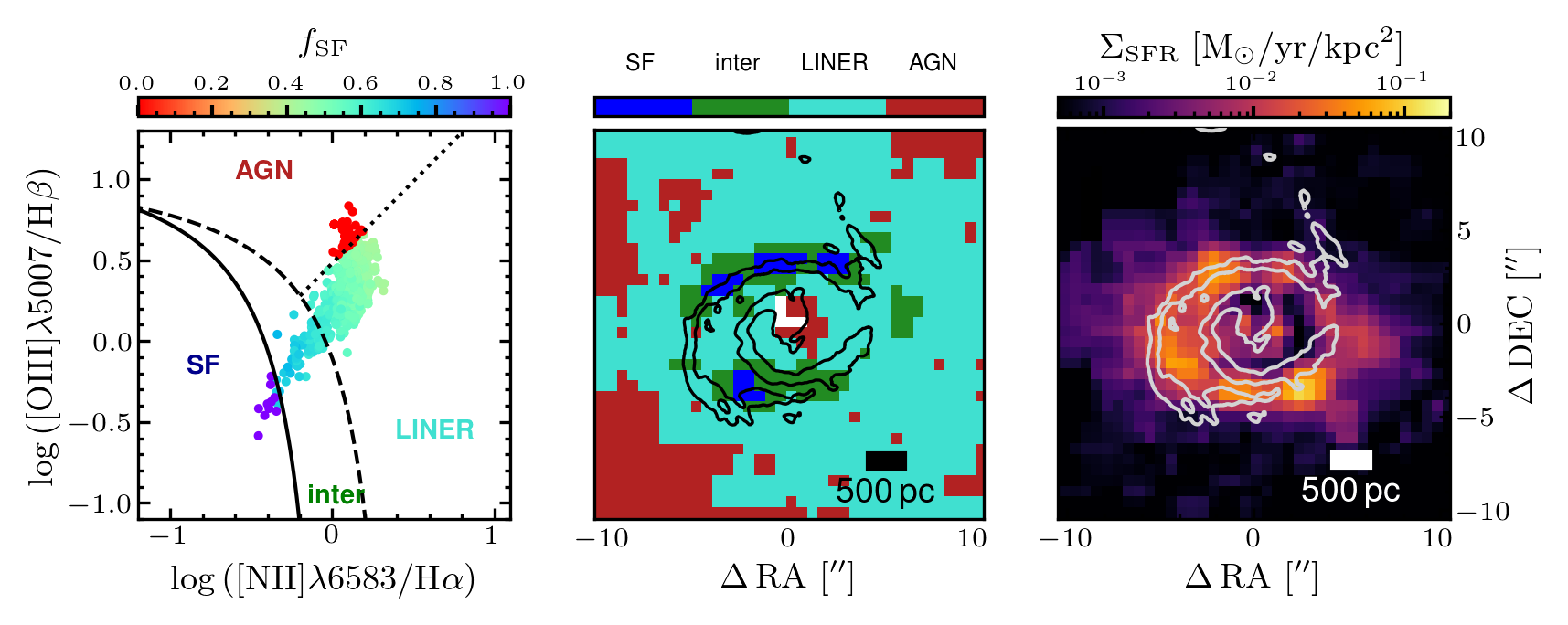}
 \caption{
 Star formation rate across the molecular gas spiral.
 The left panel shows the BPT diagnostic diagram for the combined MUSE WFM+NFM emission line maps, binned to $100\,{\rm pc}$ resolution. Ionised gas in \ngc's innermost 1.3\,kpc form a mixing sequence, with the star-forming fraction $f_{\rm SF}$ modelled using \texttt{Rainbow}. The right panel shows the resulting star formation rate surface density ($\Sigma_{\rm SFR}$), overlaid with molecular gas spiral contours. Although the spatial distribution of $\Sigma_{\rm SFR}$ roughly aligns with the molecular gas, the peaks of the two structures are offset by up to $\sim 200\,{\rm pc}$.
}
 \label{fig:BPT_SF_map}
\end{figure*}
The estimated $f_{\rm SF}$ is shown in the BPT diagram in the  left panel of Fig.~\ref{fig:BPT_SF_map}, binned to 100\,pc resolution. The spaxels within the innermost 1.3\,kpc of \ngc\ follow a continuous mixing sequence, indicating varying levels of AGN and SF excitation. Most of the field is dominated by low-ionisation narrow emission-line region (LINER)-like ratios, typically associated with diffuse ionised gas or composite AGN+SF contributions. In contrast, the emission line ratios within the spiral arm are indicative of ionisation predominantly by star formation. Very close to the nucleus ($<$1\arcsec), the hard ionisation field from the AGN -- and potentially shocks from the outflowing ionised gas \citep{Mulumba:2024} -- dominates the line emission, making the SFR estimates in this central region uncertain.

Using $f_{\rm SF}$, we corrected the H$\alpha$ emission to isolate the SF-only contribution. The resulting SFR surface density map is shown in the right panel of Fig.~\ref{fig:BPT_SF_map}. 
Finally, we adopted the SFR calibration based on extinction-corrected H$\alpha$ luminosity from \cite{Calzetti:2007}
\begin{equation}
\left( \frac{\mathrm{SFR}_{\mathrm{H}\alpha}}{[\mathrm{M}_\odot\, \mathrm{yr}^{-1}]} \right) = 5.3 \times 10^{-42} \left( \frac{L_{\mathrm{H}\alpha}}{[\mathrm{erg\, s}^{-1}]} \right).
\end{equation}

Star formation in \ngc's centre is concentrated along the molecular gas spiral, with a total star formation rate of ${\rm SFR}=(4.9 \pm 0.3) \times 10^{-2} \,{\rm M_\odot/yr}$, and  $\Sigma_{\rm SFR}$ peaking at $0.11\,{\rm M}_\odot\,{\rm kpc}^{-2}$ (see Fig.~\ref{fig:BPT_SF_map}, right panel). Notably, the peaks in $\Sigma_{\rm SFR}$ are offset from those of the molecular gas by up to $\sim200$\,pc, possibly due to spatial offsets between gas compression and subsequent star formation.

To assess the consistency of SF estimates, we compared our results to those of \citet{Diaz-Garcia:2021}, who used aperture-integrated CO(1–0) emission (beam FWHM $\theta_{\rm beam}=21\farcs5$, corresponding to 3.6\,kpc) to derive a molecular gas surface density of $\Sigma_{\rm mol}^{<3.6\,{\rm kpc}} = 38.10 \pm 1.42 \,{\rm M}_\odot/{\rm pc}^2$ and $\Sigma_{\rm SFR}^{<3.6\,{\rm kpc}} = (6.3 \pm 0.6) \times 10^{-2} \,{\rm M}_\odot/{\rm yr}/{\rm kpc^2}$ from GALEX near- and far-UV data. For the same aperture, our analysis yields $\Sigma_{\rm mol}^{3.6\,{\rm kpc}} = 31.6 \pm 2.4 \,{\rm M}_\odot/{\rm pc}^2$, consistent with their molecular gas estimate. In contrast, our $\Sigma_{\rm SFR}^{3.6\,{\rm kpc}} = (1.4 \pm 0.3) \times 10^{-2}\,{\rm M}_\odot/{\rm yr}/{\rm kpc^2}$ is slightly lower than their UV-based estimate. This discrepancy likely arises because UV fluxes from GALEX are susceptible to contamination from the AGN continuum and emission lines. Indeed, if we repeat our analysis without correcting the H$\alpha$ flux for AGN contribution, we obtain $\Sigma_{\rm SFR}^{3.6\,{\rm kpc}} = (2.3 \pm 0.5) \times 10^{-2}\,{\rm M}_\odot/{\rm yr}/{\rm kpc^2}$, bringing our results into closer agreement with \citet{Diaz-Garcia:2021}. This comparison underscores the importance of correcting for AGN contamination when estimating SFRs from emission line or continuum diagnostics in active galaxies.

\subsection{Star formation along the single-arm spiral}
\label{subsec:Discussion_SFR_Mmol}
\begin{figure}
 \centering
 \includegraphics{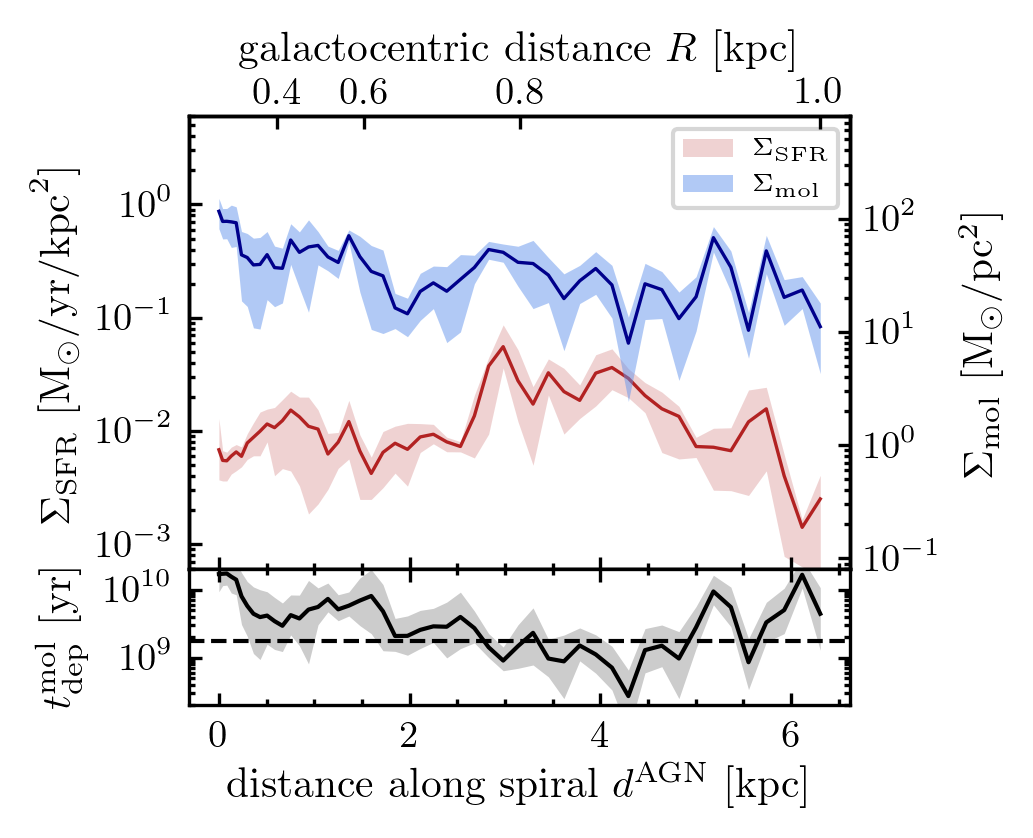}
 \caption{
 Star formation rate, molecular gas density, and depletion timescale along the single-arm spiral.
 The top panel shows the SFR surface density $\Sigma_{\rm SFR}$ (red) and the molecular gas mass surface density $\Sigma_{\rm mol}$ (blue), as function of distance to the AGN measured along the spiral arm.
 While the $\Sigma_{\rm mol}$ has high values all across the spiral, the SF is clumped, resulting in variations of $\Sigma_{\rm SFR}$ by more than one order of magnitude.
 The bottom panel shows the molecular gas mass depletion timescale, with the dashed line marking the typical $ t^{\rm mol}_{\rm dep}=1.7\,{\rm Gyr}$ measured by resolved observations in nearby spiral galaxies \citep{Utomo:2018}. In the single arm spiral, it is nearly constant with a median $\langle t^{\rm mol}_{\rm dep} \rangle = 3.9\,{\rm Gyr}$ indicating that star formation in the single-arm spiral is remarkably inefficient.
}
 \label{fig:SF_radial}
\end{figure}
Not all the molecular gas in the single-arm spiral will ultimately reach the AGN, and contribute to grow the black hole. HQ10 suggest that only a small fraction ($<$\,1\%) from the 100\,pc scale reaches the black hole accretion disc. To empirically assess the efficiency of SF in \ngc's gas inflow, we measure $\Sigma_{\rm mol}$, $\Sigma_{\rm SFR}$, and the resulting molecular gas depletion time ($t_{\rm dep} = \Sigma_{\rm mol}/\Sigma_{\rm SFR}$), along the single-arm spiral.
\begin{figure*}
 \centering
 \includegraphics[width=0.95\textwidth]{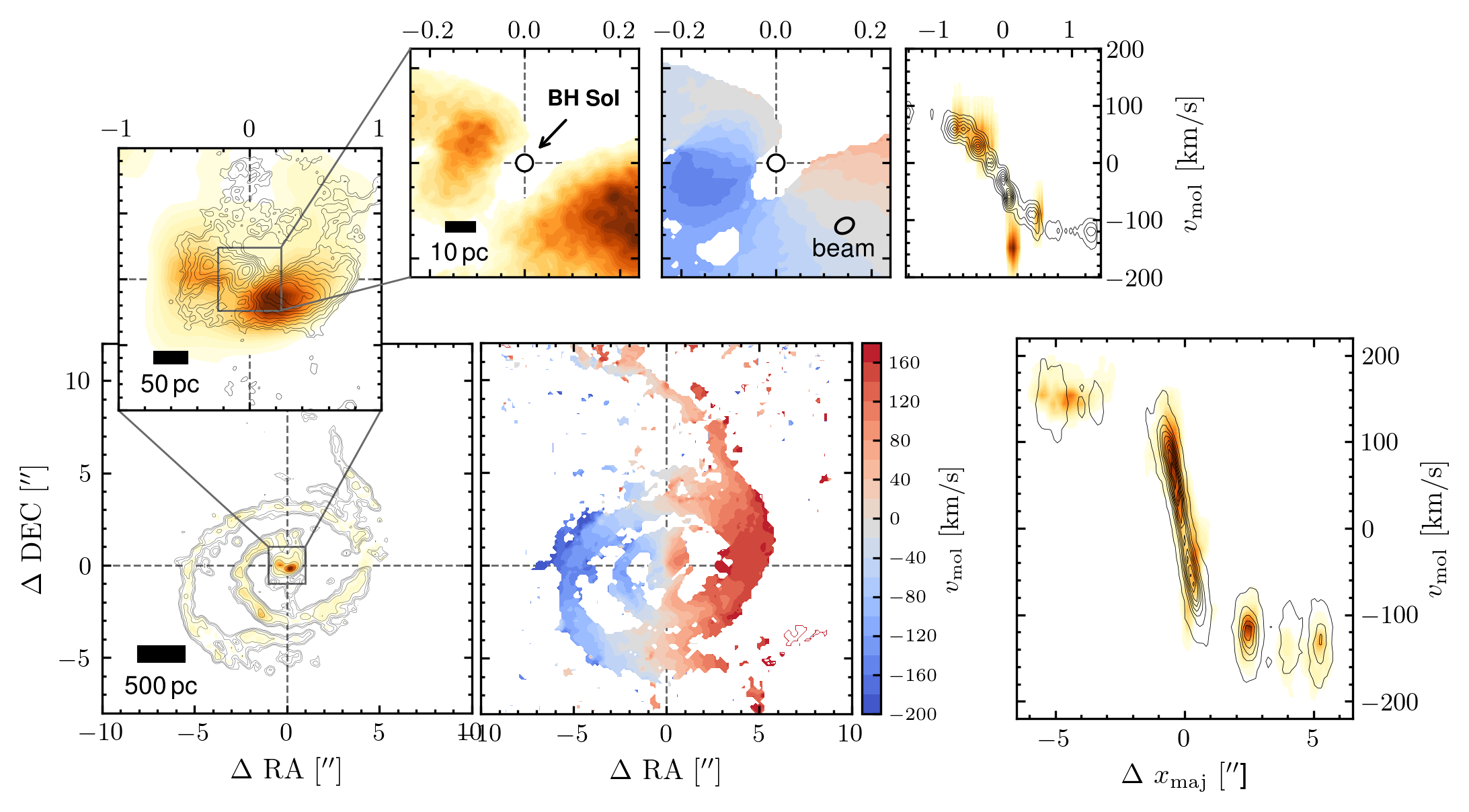}
 \caption{
     Results of the kinematic modelling carried out with \texttt{KinMS}.
     (Left) Surface brightness distribution and line-of-sight velocity field of the CO(2--1) emission, for the low- (bottom) and high-resolution (top) data set respectively. 
     Contours correspond to the best-fit model.
     (Right) Position-velocity diagram along the kinematic major axis. 
     The model includes circular rotation with radial motions, which dominate the bulk molecular gas kinematics from 1.3\,kpc down to the black hole SOI.
}
 \label{fig:KinMS_modelling}
\end{figure*}

As a benchmark, the resolved Kennicutt–Schmidt relation \cite[rKS, $\Sigma_{\rm SFR} = A(\Sigma_{M_{\rm mol}})^{\alpha}$, ][]{Kennicutt:2007, Bigiel:2008, Leroy:2008, Onodera:2010, Kreckel:2018} connects $\Sigma_{\rm mol}$ and $\Sigma_{\rm SFR}$ on sub-kiloparsec scales with a typical slope near unity and a scatter of $\sim$0.3 dex. The slope and scatter vary with spatial resolution, as noted by \cite{Sanchez:2021} and \cite{Pessa:2021}, due to the resolving of individual molecular clouds at small scales. To preserve statistical coherence while maintaining spatial detail, we resampled the emission-line maps retrieved in Sect.~\ref{subsec:star_formation_rates} to a common grid with 100\,pc\,$\times$\,100\,pc resolution. This resolution remains above the full resolution of giant molecular clouds, which are partially resolved at this scale \citep{Pessa:2021}, but is sufficient to track radial trends across the spiral.

To trace the spiral structure, we parameterized its path in polar angle $\theta$ using a second-order polynomial:
\begin{equation}
    \mathbf{r}(\theta) = 
    \begin{pmatrix}
        \Delta x \\
        \Delta y
        \end{pmatrix} = (1 - a \cdot \theta)  \theta  \cdot b \begin{pmatrix}
        \sin\,\theta \\
        \cos\,\theta
    \end{pmatrix}
\end{equation}
where $\Delta x$ and $\Delta y$ are the deprojected Cartesian distances from the AGN, and $a=0.04$, $b=-0.93$ are best-fitting parameters over the range $\theta = [0,3\pi]$. This model reproduces the spiral from 1.3\,kpc down to 100\,pc from the AGN. Along this spiral path, we extract $\Sigma_{\rm mol}$ and $\Sigma_{\rm SFR}$ in 200\,pc-wide apertures -- about three times the radial FWHM of the gas surface density profile -- to capture both the spiral’s ridge and its outskirts. This is important because $\Sigma_{\rm SFR}$ does not always align with peaks in $\Sigma_{\rm mol}$. This misalignment may indicate delayed star formation along the trailing edges of the spiral arm, as observed in other galaxies \citep{Chandar:2017, Williams:2022}.

We find a median $\Sigma_{\rm SFR} = (9.6 \pm 1.3) \times 10^{-3} \,{\rm M}_\odot/{\rm yr}/{\rm kpc^2}$ with uncertainty reflecting spatial variability. The median molecular gas surface density is $37 \pm 3\,{\rm M}_\odot/{\rm pc}^2$ which corresponds, via the 100\,pc rKS relation from  \cite{Pessa:2021} ($A=-9.96$, $\alpha_{100\,{\rm pc}} = 1.06$), to an expected SFR of $1.3 \times 10^{-2}\, {\rm M}_\odot/{\rm yr}$ -- larger than our measurement. This implies moderate star formation efficiency along the spiral. Associated depletion timescales tend to be long, but vary substantially, as shown in the bottom panel of Fig.~\ref{fig:SF_radial}. Radial positions are expressed as the deprojected distance to the AGN ($d^{\rm AGN}$), with a secondary axis showing the galactocentric radius ($R$). The median molecular gas depletion time, $\langle t^{\rm mol}_{\rm dep} \rangle =  3.9 \pm 0.6\,$\,Gyr, is significantly larger than the typical $\tau_{\rm dep} \sim 1$–$2$\,Gyr observed in the central regions of nearby low-luminosity AGNs \citep{Casasola:2015} and in the spiral arms of disc galaxies \citep{Leroy:2013}, indicating inefficient star formation along the spiral.

All derived quantities along the spiral are based on aperture-integrated fluxes. Aperture sizes were chosen to avoid resolving individual clouds, though we note that the cloud scale can vary significantly. Nonetheless, the qualitative trends of $\Sigma_{\rm mol}$, $\Sigma_{\rm SFR}$, and $t_{\rm dep}$ are robust resolution changes between 100 and 300\,pc, and across aperture widths ranging from 100 to 400\,pc.

\subsection{Molecular gas kinematic modelling}
\label{subsec:kinematic_modelling_mol_gas}
\begin{table}
\caption{Results of modelling the molecular gas kinematics with \texttt{KinMS}.}
\label{tbl:KinMS_results}     
\centering
\begin{small}

\begin{tabular}{lcccc}
  \hline\hline 
  \noalign{\smallskip} 
    Parameter &  
    Initial guess  \tablefootmark{(a)} & 
    Best-fit outer \tablefootmark{(b)} & 
    Best-fit inner \tablefootmark{(c)} \\ [0.5ex] 
  \hline 
  \noalign{\smallskip} 
    $F$ \:[K\,km\,s$^{-1}$]                            & 25     & $189.2^{+4.2}_{-9.2}$   & $66.1 ^{+0.3}_{-1.2}$           \\ 
    PA \:[$^\circ$]                           & 290    & $277.6 ^{+1.1}_{-2.1}$  & $ 279.4^{+0.8}_{-0.3}$     \\ 
    $i$ \:[$^\circ$]                           & 10     & $48.8^{+3.9}_{-3.5}$   &   $36.0^{+0.5}_{-0.6}$          \\ 
    $ x_0$ \:[$\arcsec$]                              & 0      & $0.02^{+0.02}_{-0.03}$ & $-0.03^{+0.04}_{-0.02}$   \\  
    $ y_0$ \:[$\arcsec$]                              & 0      & $0.03^{+0.02}_{-0.02}$ & $-0.04^{+0.03}_{-0.02}$   \\  
    $\Delta v_{\rm sys}$ [km\,s$^{-1}$]                & 2450   & $2487.5^{+1.1}_{-1.9}$ & $2484.6^{+1.8}_{-0.9}$            \\ 
    $\sigma_{\rm gas}$ [km\,s$^{-1}$]                  & 20     & $20.81^{+1.40}_{-1.41}$ & $31.2^{+0.2}_{-0.3}$         \\ 
    
    $ R_{\rm turn}$\: [$\arcsec$]              & 1.0     & $1.06^{+0.25} _{-0.12}$ &  1.06 (fixed)        \\ 
    $ v_{\rm max}$ \:[km\,s$^{-1}$]                    & 200    & $232.4^{+4.9}_{-11.7}$ & $218.7^{+4.5}_{-8.7}$        \\ 
    ${\rm log}\, M_{\rm BH}/{\rm M}_\odot$             & 6.65    &    6.65 (fixed)                 & 6.65 (fixed)       \\ 
\noalign{\smallskip}\hline
\hline
\end{tabular}
\tablefoot{
\tablefoottext{a}{Initial guess for the parameters.} 
\tablefoottext{b}{Best-fitting value retrieved from modelling the 15\arcsec\,$\times$\,15\arcsec\ lower-resolution cube.} 
\tablefoottext{c}{Best-fitting value retrieved from modelling the inner 1\arcsec\,$\times$\,1\arcsec\ of the high-resolution cube.} 
}                              

\end{small}
\end{table}

To constrain the kinematics of the molecular gas, we employ the Kinematic Molecular Simulation (KinMS) routines from \cite{Davis:2013}, specifically the \texttt{KinMS\_fitter}\footnote{\url{https://www.kinms.space}}. This forward-modelling approach infers the kinematic and dynamical parameters of the molecular gas distribution in interferometric datacubes. To account for the asymmetric flux distribution of the spiral, we use an intensity-weighted sampling generated by the \texttt{skySampler} plugin for \texttt{KinMS\_fitter}. The overall motion is dominated by disc-like rotation. We set up the model with an arctan-rotation curve of the form $v(R) = 2/ \pi \times v_{\rm max} \times \arctan(R/R_{\rm turn})$, yielding ten free parameters: The total CO flux $F$, the maximum velocity $v_{\rm max}$, turnover radius $R_{\rm turn}$, and the position angle PA, inclination of the disc $i$, and offsets $\Delta x_0$, $\Delta y_0$, and $\Delta v_{\rm sys}$ relative to the assumed dynamical centre and systemic velocity, respectively. For the high-resolution data cube, we also include the black hole mass \mbh to account for increasing circular velocities at very small radii. Additionally, we include pure radial motions $v_{\rm rad}$ which vary freely in 20 bins of galactocentric distance $R$.

Although we combined the medium- and high-resolution data sets into a single data cube, we model the kiloparsec-scale spiral and the inner $\sim$100\,pc region independently. This approach reduces MCMC runtime while ensuring that non-circular motions on all scales are properly accounted for. The `outer' spiral is modelled using the full cube at a resolution of $b_{\rm maj} = 0\farcs23$, while the `inner' region is modelled using a $1 \farcs 1 \times 1 \farcs 1$ cutout cube at a resolution of $b_{\rm maj} = 40\,{\rm mas}$.

We first performed a simple fit on the `outer' cube to obtain starting parameters for refinement. These starting parameters, listed in Table~\ref{tbl:KinMS_results}, were then applied to both cubes. For the MCMC run, we assumed uniform priors with sensible boundaries, and ran 30,000 samples. The best-fit \text{KinMS} models for \ngc's surface brightness and velocity fields are shown Fig.~\ref{fig:KinMS_modelling}. Disc-like rotation dominates the kinematics on across the 2.5 windings of the single-arm spiral, spanning $\sim$1.3\,kpc. Within the innermost 35\,pc, however, a fast-moving bright component -- possibly a molecular gas outflow -- dominates the surface brightness, though near-disc-like rotation remains traceable down to the black hole SOI. This outflow feature is distinct from the one identified by \cite{Kianfar:2024}, who observe a non-axisymmetric feature north-east of the nucleus. In contrast, the fast-moving gas we identify is south-west of the nucleus and shows a distinct emission component in the PV diagram (Fig.~\ref{fig:KinMS_modelling}, top right), inconsistent with circular rotation. Regardless, the underlying disc component is well-described by the \texttt{KinMS} model, and as such, this feature is not further analysed and is excluded from subsequent discussions. Although low- and high-resolution data sets were fitted independently, the best-fitting kinematic parameters, listed in Table~\ref{tbl:KinMS_results}, are consistent with each other. This consistency suggests that a single velocity profile is sufficient to describe the molecular gas kinematics across three orders of magnitude in spatial scale, from 1.3\,kpc down to 3.4\,pc from the black hole SOI (1.7\,pc), the closest radius where CO is detected.

Using the best-fit kinematic model, we estimate radial mass transport rates by evaluating $v_{\rm rad}$ in bins of radial distance along the semi-major axis. For each radial bin $i$, $v_{\rm rad}$ is weighted by the molecular gas mass within the bin and divided by the bin's radial size, $d_i$, to estimate the mass inflow rate as \mbox{$\dot{M}_{\rm mol} = M_{\rm mol} \, v_{\rm rad} / d_i$}.  Indeed, we measure $v_{\rm rad}<0$ at all radii, which can be interpreted as mass inflow along the spiral. However, kinematically measured radial components should not be directly equated with inflow or outflow, as they may also result from stable elliptical orbits. The radial motions and potential mass flow rates are further discussed in Sect.~\ref{subsec:Torques_Young_Stars}, where we also compare them with the mass inflow rates derived from gravitational torques.

\subsection{Extended UV emission}
\label{subsec:UV_photometry}
\begin{figure*}
 \centering
 \includegraphics[width=\textwidth]{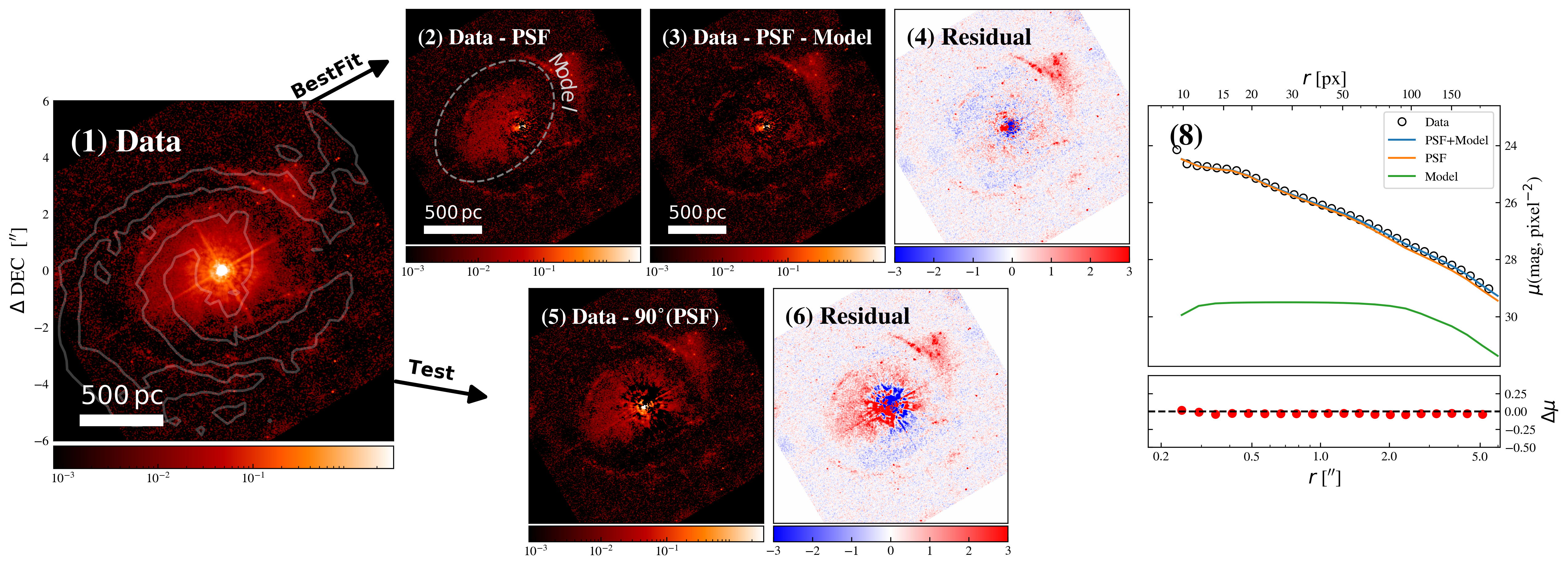}
 \caption{
    Photometric decomposition results using \texttt{lenstronomy}. 
    (1) HST/ACS HRC F330W flux map with saturated pixels replaced. Grey lines show surface brightness contours of $\Sigma_{\rm mol}$.
    (2) F330W flux map after subtracting the PSF model, revealing a diffuse component near the centre. Dashed contours show the initial guess of the single-Sérsic model fit to this structure.
    (3) F330W residual image after subtracting both the PSF and best-fit single-Sérsic model.
    (4) Residual map normalized by uncertainty.
    (5) Same as (3) but subtracting a 90$^\circ$-rotated PSF from the original image.
    (6) Similar to (4) but the single-Sérsic model was not subtracted.
    The detection of the diffuse component near the AGN does not depend on the PSF subtraction (2 vs 5). 
    Aside from compact UV emission from star-forming clumps in the single-arm spiral, no extended UV emission is left over when a single-Sérsic model is subtracted.
}
 \label{fig:lenstronomy_final_qso_fit}
\end{figure*}

\begin{figure}
 \centering
 \includegraphics[width=0.85\columnwidth]{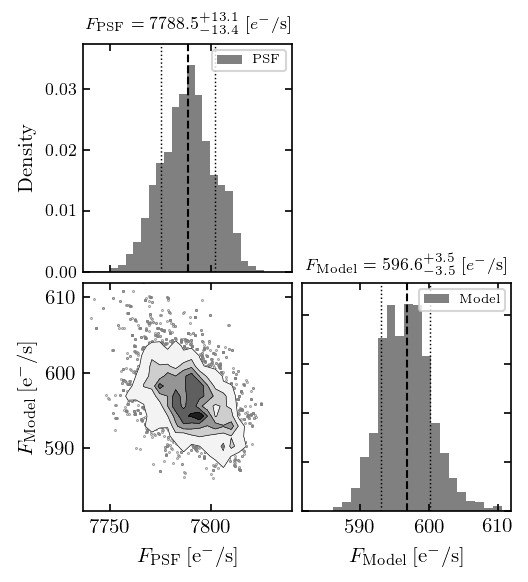}
 \caption{
    Posterior distribution between the flux of two components as recovered with \texttt{lenstronomy}.
    The bottom-left panel shows the covariance between the PSF flux and the flux of the Sérsic model fit to the diffuse off-centre component detected in UV emission. The two flux measurements are loosely correlated, and the Sérsic component is detected with high significance.
}
 \label{fig:lenstronomy_flux_corner}
\end{figure}

Although the current star formation rate in the single-arm spiral is modest, the abundant molecular gas provides the fuel for young, UV-bright stellar populations in the centre of \ngc. To identify and characterize this component, we analysed the high-resolution HST/ACS F330W archival image, performing a photometric decomposition to determine the location, mass, and morphology of the young stars.

\subsubsection{Photometric decomposition}

Accurate decomposition requires a high-quality model of the PSF. However, PSF models generated by \texttt{Tiny Tim} \citep{Krist&Hook:2011} for this instrument-filter combination suffer from systematic limitations, such as mismatches in resolution and field coverage. Furthermore, the FOV of \ngc contains no bright field stars suitable for empirical PSF construction. We therefore selected a library of standard stars observed under similar conditions, chosen for their proximity to \ngc's AGN in FoV alignment and exposure settings. Specifically, we used the white dwarfs GD\,71, GD\,153, HD\,229196, and HD\,46106, as their exposures approach the saturation limit and their spectral energy distributions are closest to that of an AGN.

For the photometric decomposition, we employed the open-source software \texttt{galight} \footnote{\url{https://github.com/dartoon/galight}} \citep{Ding:2020}, which uses the two-dimensional image modelling provided by \texttt{lenstronomy} \citep{Birrer:2021}. To create a PSF with an S/N comparable to that of the AGN, we used \texttt{galight} to stack one, three or five library PSFs to subtract scaled versions from the \ngc image. As shown in Figure~\ref{fig:lenstronomy_final_qso_fit}, several extended UV features persists after PSF subtraction. In addition to the compact UV emitting regions co-spatial with the molecular gas, we detect a diffuse component located $\sim 1\arcsec $ east of the galaxy centre.

To investigate the nature of this component, we modelled its surface brightness distribution using \texttt{galight} with a single-Sérsic profile. Although this parameterization has no deeper physical motivation, it is sufficiently flexible to reproduce the surface profile and total, allowing reliable measurements of the integrated luminosity and spatial extent. Uniform priors were applied to the effective radius $ 0\farcs01 < R_{\rm eff} <  4\farcs0$ and Sérsic index $ 0.3 < n_{\mathrm{S\acute{e}rsic}}< 9$, ensuring reasonable constraints on the inferred parameters. The initial position of the Sérsic component was set to approximately [-30 pixels, -35 pixels], with a position angle of -65$^{\circ}$. To isolate the diffuse component, we masked the compact UV emission from star-forming regions in the spiral arms $\sim 1.5 \arcsec$ north-west of the AGN, as these regions are not the primary focus of this analysis. In order to reduce systematic uncertainties, we also masked the innermost $0\farcs2$ of the image, where the diffuse component overlaps with AGN residuals. The best-fit parameters and nominal uncertainties derived from \texttt{galight}'s Markov chain Monte Carlo decomposition are summarized in Table~\ref{tbl:lenstronomy_results}. The corner plot of absolute fluxes in Fig.~\ref{fig:lenstronomy_flux_corner} shows that the PSF flux and the Sérsic component flux uncertainties are only weakly correlated. The covariance is not strong, because the Sérsic component's offset from the centre is significant (1\farcs51 or 258\,pc), a displacement comparable to its semi-major axis half-light radius, $R_{\rm S\acute{e}rsic}$.

To determine whether the extended feature is real or an artefact, we tested its sensitivity to systematics from PSF subtraction. Regardless of the specific PSF model, whether using a subset of standard stars or rotating the by 90\degr\ before subtraction (panels 5 and 6 of Fig.~\ref{fig:lenstronomy_final_qso_fit}), the extended feature near the centre persists. This rules out residual structure arising from azimuthal PSF asymmetries. A possible source of UV flux near \ngc's centre could be UV photons from the AGN continuum emission scattering off ambient dust and free electrons in the ISM. This effect has been observed in Seyfert galaxies on 100\,pc - 1\,kpc scales \citep[e.g.][]{Neff:1994} and while it can contribute to the total UV emission, its overall contribution to extended UV emission is often small ($\lesssim10\%$) \citep{Munoz-Marin:2009}. Alternatively, emission lines from ionized gas -- specifically [\ion{Ne}{V}]$\lambda\lambda$3346,3426, the only emission lines that significantly affect the F330W filter at \ngc's redshift -- could contribute. This would be accompanied by [\ion{O}{iii}] emission on similar scales \citep{Munoz-Marin:2009}, but it is remarkably weak at the location of the diffuse UV component (see Sect.~\ref{subsec:star_formation_rates}). We therefore favour a third scenario: the diffuse UV emission in \ngc's centre originates from an unresolved young stellar population. Its diffuse nature may result from a population of smaller star clusters or the disruption of ageing clusters \citep{Colina:1997, Fanelli:1997, Munoz-Marin:2009}.  However, it is possible that residuals in the UV morphology modelling, especially those close to the AGN (see panel 4 Fig.~\ref{fig:lenstronomy_final_qso_fit}), have a contribution from an overly simplistic model or scattered photons from the AGN.

\begin{table}
\caption{Results of modelling the diffuse UV component with \texttt{galight}.}
\label{tbl:lenstronomy_results}      
\centering
\begin{small}

\begin{tabular}{lcccc}
  \hline\hline 
  \noalign{\smallskip} 
    Parameter &  
    Best-fit value\\
  \hline 
  \noalign{\smallskip} 
   Amplitude \:[e$^{-}/{\rm s}$]  &  $ 19.73_{-0.30}^{+0.24}	$ \\ 
   $R_{\rm S\acute{e}rsic}$ [$^{\prime\prime}$]&  $  1.47_{-0.02}^{+0.02}	 $ \\ 
   $n_{\rm S\acute{e}rsic}$ &  $0.31_{-0.01}^{+0.01}$ \\ 
   $x_c$ [$^{\prime\prime}$]  &  $   1.00_{-0.02}^{+0.03}	$ \\ 
   $y_c$ [$^{\prime\prime}$]  &  $  -1.13_{-0.01}^{+0.02} $ \\ 
   $q$ &  $  0.67_{-0.02}^{+0.01} $ \\ 
   PA [$^{\circ}$]&  $ 75.5_{-2.0}^{+2.5}	$ \\ 
   $F ({\rm F330W}) $\: [erg/s] &  $ 224_{-4}^{+3}$ \\ 
   $m_{\rm AB}({\rm F330W})$ [mag] &  $ 18.20_{-0.01}^{+0.02}	 $ \\
\noalign{\smallskip}\hline
\hline
\end{tabular}          
\end{small}
\end{table}

\subsubsection{Mass of the young stellar component}
\label{subsec:characterizing_young_component}
To determine whether this off-centred young stellar component is cause or result of the single-arm gas spiral, we aim to constrain its stellar mass, $M_\star$. While an archival HST/WFC3 F547M image exists, dust obscuration does not allow for a photometric decomposition on similar scales. Thus, we were limited to a flux measurement in the F330W filter. While this measurement only permits estimating few parameters from the spectral energy distribution, it suffices to constrain lower and upper bounds for $M_\star$ by adopting sensible estimates for the properties of the underlying stellar populations. For this task, we employed \texttt{pygalaxev}\footnote{\url{https://github.com/astrosonnen/pygalaxev}} to create composite stellar population models and predict the F330W broadband magnitudes on a grid of stellar population parameters. Specifically, we used the stellar population models generated with the \texttt{GALAXEV} code \citep[][2016 version CB16]{Bruzual:2003}.

We interpret the diffuse UV component as originating from young stars formed out of the inflowing molecular gas in the single-arm spiral. The youngest ($<$5\,Myr) stellar populations appear as UV and H$\alpha$-emitting clumps in the outer ($>$300\,pc) spiral arm are visibly compact ($\lesssim $50\,pc diameter), e.g. north-west of the nucleus (see Fig.~\ref{fig:lenstronomy_final_qso_fit} left panel). In contrast, the central diffuse UV emission, with a half-light radius of $R_{\rm S\acute{e}rsic} = 1\farcs47$ (253\,pc), lacks the compact morphology typical of OB associations. This suggests it arises from a somewhat older stellar population (10–200\,Myr), whose initial clustering has dispersed through dynamical processes.

We therefore assumed that molecular gas forming the young stars to be pre-enriched with metals. This is motivated by the observation that the inner regions of the galaxy bar's dust lanes -- which, in the barred potential, channel gas from the galaxy's outskirts to the inner $\sim$2\,kpc \citep{Sormani:2015} -- are directly connected to the outer ends of \ngc's nuclear single-arm spiral. This connection implies that the gas originates from \ngc's outer star-forming disc and has already traversed the entire galaxy before reaching the nuclear single-arm spiral. Consequently, we assume a metallicity range of 0.01 < Z/Z$_\odot < 5$. Emission line maps (Sect.~\ref{subsec:star_formation_rates}) within a 1\farcs5 aperture around the young stellar component show substantial variations in extinction (when detected), leading us to adopt a range for the optical depth of $0.1 < \tau_V < 3$. Finally, we assume an exponential decay timescale for the SFR of 0.1\,Gyr $< \tau$ $<$ 2\,Gyr. Within these parameter ranges, the stellar mass of the young component is constrained to $10^{7.4} < M_\star/ {\rm M}_\odot < 10^{9.3}$.

\subsection{Gas inflow driving mechanism}
\label{subsec:Driving_Mechanism}
A single-arm morphology in the molecular gas is characteristic of the $m=1$ mode, analogous to a dipole-like gravitational potential \citep{Jog:1997, Jog&Combes:2009}. Such structures can arise from tidal encounters \citep{Zaritsky&Rix:1997} or counter-rotating discs \citep{Comins:1997, Dury:2008}, but no such interactions are observed in \ngc. Instead, we propose that the mass of the young stars, formed from the inflowing gas and offset from the galaxy centre, creates a one-sided excess in stellar mass density. This lopsided mass distribution acts as a gravitational perturber, driving the single-arm pattern in the inflowing gas. In the following, we derive the gravitational torque exerted by the asymmetric young stellar population on the molecular gas.

\subsubsection{Torques: Methodology}
\label{subsubsec:torques_methodology}
\begin{figure}
 \centering
 \includegraphics{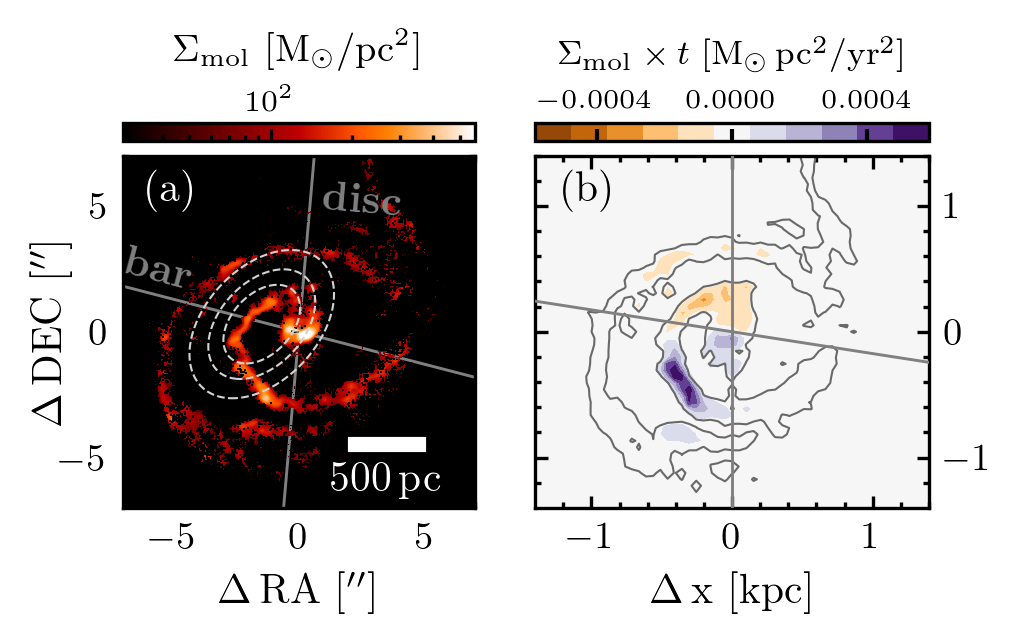}
     \caption{
        Torque acting from \ngc's diffuse young stellar component onto the molecular gas spiral.
        The left panel shows the molecular gas mass surface density of the single-arm spiral together with the contours of the mass distribution of the young stellar component. The right panel shows a de-projected map of the torques acting onto the molecular gas.
    }
 \label{fig:mass_torque_maps}
\end{figure}
To quantify the torques, we followed the method of \citet{Garcia-Burillo:2005}, which is commonly used to estimate $\dot{M}_{\rm gas}$ in galaxy discs via gravity torques and angular momentum transport \citep[e.g.][]{Haan:2009, Querejeta:2016}. As a first step, we transformed the coordinates to the galaxy’s face-on plane. We assume that the molecular gas lies on near-circular orbits and spatially distributed in a thin disc with negligible height, as it is typical for disc galaxies \citep{Jeffreson:2022}. The young stellar component is modelled as an isothermal plane with a vertical scale height of approximately one-twelfth of the radial scale length. We deprojected the coordinates using an affine transformation with a position angle ${\rm PA} = 5^\circ$ and inclination $i=38^\circ$, based on the gas disc parameters described in Sect.~\ref{subsec:kinematic_modelling_mol_gas}. From the resulting deprojected mass surface density, we compute the gravitational potential $\Phi(x,y)$ and the corresponding gravitational force per unit mass $\Vec{F}(x,y)$ at each pixel, which was use to derive the specific torque as
\begin{equation}
    \label{eq:torques}
    t(x,y) = xF_y -yF_x\,. 
    \end{equation}
The torque map across \ngc's single-arm spiral is shown in the right panel of Fig.~\ref{fig:mass_torque_maps}, where negative torques indicate angular momentum loss in the rotating gas. We estimate the time derivative of the angular momentum surface density $dL_s(x,y)/dt$ by weighting $t(x,y)$ with the molecular gas column density $\Sigma_{\rm mol}(x,y)$ as derived from the CO(2-1) line maps. We then derived the radial profile of the specific angular momentum loss by azimuthally averaging the torque per unit mass weighted by the molecular gas surface density $\Sigma_{\rm mol}(x,y)$
\begin{equation}
    \label{eq:angualar_momentum_loss_rate}
    t(R) = \frac{\int_\theta \Sigma_{\rm mol}(x, y) \times \left( xF_y - yF_x \right)}{ \int_\theta \Sigma_{\rm mol}(x, y) }\,.
\end{equation}
We also adopted the definition of the dimensionless AGN feeding efficiency $\Delta L/L$, which GB05 define as the fractional change in specific angular momentum during one orbital period, $T_{\rm rot}$):
\begin{equation}
    \label{eq:AGN_feeding_efficiency}
    \frac{\Delta L}{L} = \frac{{dL}}{{dt}} \bigg|_{\theta} \times  \frac{{1}}{{L}} \bigg|_{\theta}  T_{\rm rot} 
    = \frac{t(R)}{L_\theta} \times T_{\rm rot}\,.
\end{equation}
Under this definition, the molecular gas angular momentum $L_\theta = R \times v_{\rm rot}$ is removed entirely during one orbital period if $\Delta L/L = -1$. Finally, the radial gas mass inflow rate per unit length can be estimated from the angular momentum loss rate as:
\begin{equation}
    \label{eq:mass_inflow_rate_differential}
    \frac{d^2M}{drdt} = \frac{{dL}}{{dt}} \bigg|_{\theta} \times \frac{{1}}{{L}} \bigg|_{\theta} \times 2\pi R \times \Sigma_{\rm mol}(x, y) |_{\theta}\,,
\end{equation}
where $\Sigma_{\rm mol}(x, y) |_{\theta}$ is the radial profile of the azimuthally averaged gas mass surface density. Multiplying with the radial shell width $\Delta R$ yields the azimuthally averaged local gas mass inflow rate
\begin{equation}
    \label{eq:mass_inflow_rate}
    \dot{M}(R)= \sum \frac{{d^2M}}{{drdt}} \times \Delta R\,.
\end{equation}

\subsubsection{Torques from the young stellar component}
\label{subsec:Torques_Young_Stars}
\begin{figure}
 \centering
 \includegraphics{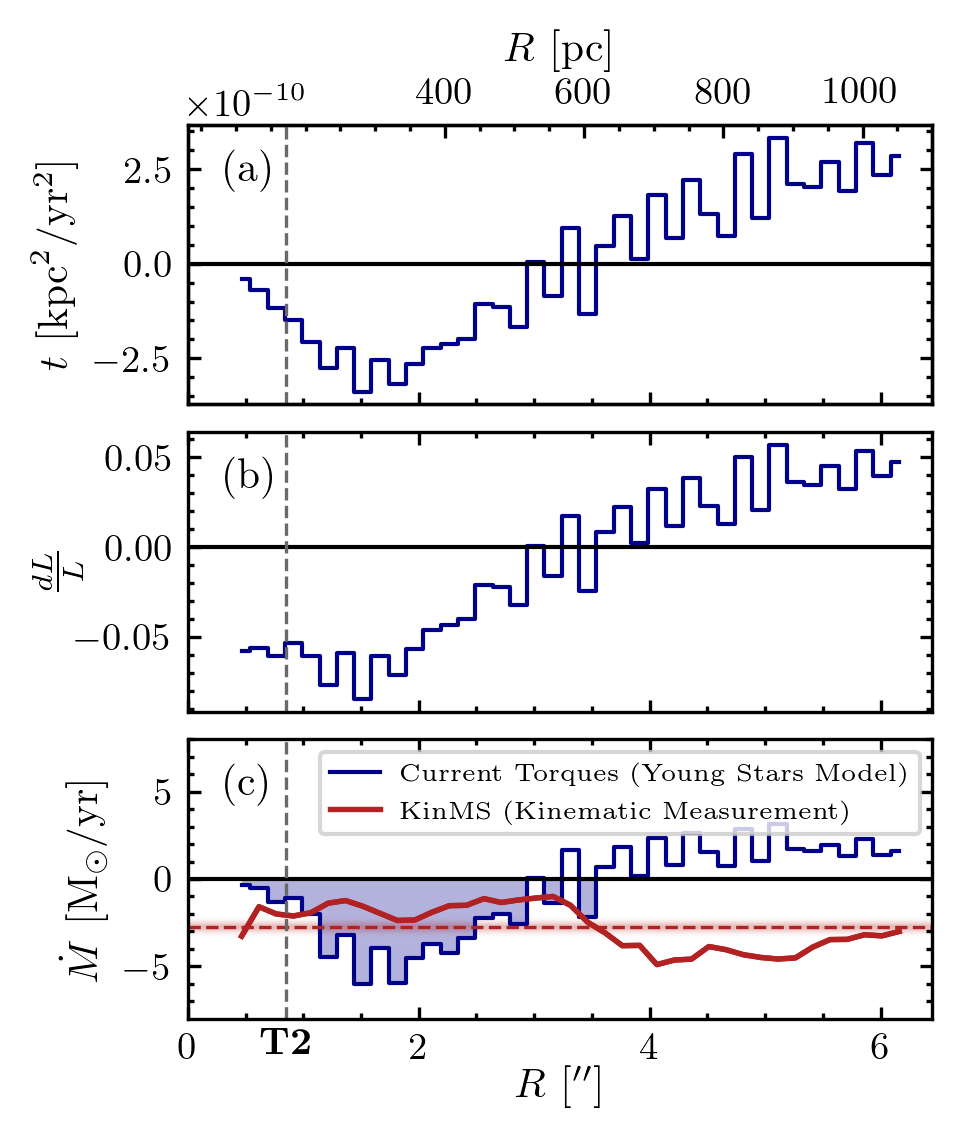}
 \caption{
 Snapshot torque budget and mass-inflow rates across \ngc's molecular gas spiral.
From top to bottom, the panels show the radial behaviour of the
(a) gravitational torques per unit mass – $t$ 
(b) the average fraction of the angular momentum transferred to the gas in one orbital period $\Delta L/L$, and
(c) the molecular gas mass inflow rate derived from stellar torques (blue line) compared to the kinematic measurement from \texttt{KinMS} (red line).
While the sign of angular momentum transfer depends on the torquing mass' orientation relative to the gas spiral, the torque amplitude is sufficient to explain the kinematically measured mass inflow rate of $\langle \dot{M}(R) \rangle = -4.2 \,{\rm M}_\odot / {\rm yr}$ (see Sect.~\ref{subsec:kinematic_modelling_mol_gas}).
}
 \label{fig:radial_torque_inflows}
\end{figure}

The young stellar component has a non-axisymmetric morphology, with its centroid displaced 258\,pc from the galaxy centre. This suggests that the young stars, formed from inflowing gas, may also funnel molecular gas towards the centre. We aim to assess whether the gravitational torque from the young stars drives the radial gas flow by computing the torques exclusively from this component, assuming it is superposed on the old stellar body, which does not significantly contribute to the overall torque budget. This approach differs from previous studies, which typically used the old stellar component mapped through near-infrared imaging to derive torques.

The morphological parameters $x_0$, $y_0$, $R_{\mathrm{S\acute{e}rsic}}$, $n_{\mathrm{S\acute{e}rsic}}$, and $q$ were taken from Table~\ref{tbl:lenstronomy_results}. The setup is illustrated in the left panel of Fig.~\ref{fig:mass_torque_maps}, which shows the surface mass density of the lopsided young stellar component clearly offset from the AGN position. The corresponding radial profiles of the azimuthally averaged specific torques, feeding efficiency and gas mass inflow rate are shown in Fig.~\ref{fig:mass_torque_maps}. We note that apart from the relatively broad constraints estimated from the UV photometry (Sect.~\ref{subsec:UV_photometry}, the mass of the young stellar component, $M_{\star,\mathrm{young}}$, remains a free parameter. For illustrative purposes, we consider $M_{\star,\mathrm{young}} = 10^{8.5}$, a value close to the median of the range discussed in Sect.~\ref{subsec:characterizing_young_component}. $M_{\star,\mathrm{young}}$ primarily scales the amplitude of $t$, $dL/L$, and $\dot{M}$ by a constant factor, without affecting the sign of the net torque $t$ (and thus $dL/L$ or $\dot{M}$) at a given radius. This means that the direction of the gas mass flow is independent of $M_{\star,\mathrm{young}}$, while its amplitude depends on it.

The behaviour of $t$ strongly depends on the relative orientation of $M_{\star,\mathrm{young}}$ and $M_{\rm mol}$, which are likely unstable. As a result, the current torques may not represent long-term averaged quantities such as $dL/L$ and $\dot{M}$. Dedicated simulations, beyond the scope of this work, could provide more reliable insights into the dynamics of the system. Within $\sim$500\,pc, these torques consistently remove angular momentum, although with a low rate, with $dL/L \sim -0.05$. When weighted by $\dot{M}_{\rm mol}$, the resulting mass inflow rates peak locally at $-7 \, {\rm M}_\odot \, \mathrm{yr}^{-1}$ and average $-4.2 \, {\rm M}_\odot \,\mathrm{yr}^{-1}$. 

To put this into context, we estimate the mass inflow rates from the molecular gas kinematics using the radial velocity component measured with \texttt{KinMS} (see Sect.~\ref{fig:KinMS_modelling}). It is important to note that the detection of radial motions does not necessarily indicate inflows or outflows, as they can also arise from stable elliptical orbits, commonly seen in galaxies. However, in \ngc the measured radial velocities are consistently negative (i.e. inward-directed) across all radii, and remain so across two-and-a-half windings of the single-arm spiral. This behaviour contrasts with the expectation from elliptical orbits, where $v_{\rm rad}$ would typically change sign twice within a single phase. Therefore, we interpret the persistently negative $v_{\rm rad}$ -- and the associated kinematic mass inflow rates ($\dot{M}_{\rm mol}^{kin}$) -- as evidence for a genuine mass inflow.
The bottom panel of Fig.~\ref{fig:radial_torque_inflows} shows that the amplitude of the kinematically measured $\dot{M}_{\rm mol}$ is broadly consistent with the theoretically derived from the torques exerted by the young stellar component (Sect.~\ref{subsec:kinematic_modelling_mol_gas}).
We note that although the kinematically and torque-derived $\dot{M}_{\rm mol}$ values agree, the high sensitivity of the torque calculation to the configuration of the young stellar component leads us to interpret this consistency with caution.
At galactocentric distances larger than $\sim$500\,pc, the torques acting on the molecular gas become positive (see top panel of Fig.~\ref{fig:radial_torque_inflows}), indicating that the gas gains angular momentum locally, which reverses the direction of the mass flow and drives the gas outward.

However, the torque-derived mass inflow rates must be interpreted with caution for two key reasons.
First, the configuration is likely unstable over time. For instance, rotating $M_{\star,\mathrm{young}}$ by 180$^\circ$ around the galaxy centre approximately reverses the sign of the radial torque budget. 
Given that the orbital period of $M_{\star,\mathrm{young}}$ is approximately three times shorter than that of the outer edges of the spiral arm, the net torque acting on the outskirts would repeatedly change sign as the gas flows inward, introducing significant time dependence on the torque budget.
Second, the torque method remains unvalidated by simulations and could yield estimates that deviate by several orders of magnitude. The method calculates inflow rates based only on the mass of individual clouds and their positions relative to the underlying gravitational potential, assigning non-zero inflow rates to clouds in non-axisymmetric potentials, even if they follow closed, eccentric orbits with zero net mass inflow.
Given these limitations, the values for $t$, $dL/L$, and $\dot{M}$ derived from the current configuration should be regarded as a "snapshot" of the system rather than as fixed quantities. Despite these caveats, the amplitude of the torques is, in principle, sufficient to account for the kinematically estimated mass inflow rates.

\section{Discussion}
\label{sec:Discussion}

\subsection{The young stellar component as angular momentum sink}
The molecular gas at the outer end of NGC 4593's single-arm spiral must lose 99.6\% of its angular momentum to reach the black hole SOI. While Sect.~\ref{Sec:Complementary_processes} explores the processes that trigger the $m=1$ instability, a key question remains as to where the angular momentum is transferred.

A sustained and continuous loss of angular momentum is necessary to drive the gas inflow, shaping the morphology of the single-arm spiral considering that it extends over 2.5 windings from 1.3\,kpc down to the black hole SOI. One possible mechanism for angular momentum removal are fast-moving outflows. Notably, high-velocity molecular gas observed at 35\,pc from \ngc's nucleus suggests the presence of a molecular outflow (Sect.~\ref{subsec:kinematic_modelling_mol_gas}).
Outflows in \ngc's centre have also been detected in the ionised gas phase, with a projected size of 340\,pc, carrying $8 \times 10^5\,{\rm M}_\odot$ at $v_{\rm max} \sim 200 \,{\rm km\,s}^{-1}$, consistent with expectations for AGNs of this luminosity \citep{Mulumba:2024}. However, both outflows are confined to <100\,pc scales, and the angular momentum carried by these outflows is negligible compared to the total required for the gas to reach the vicinity of the black hole. 
This suggests that outflows alone cannot account for the angular momentum loss over the larger distances involved. A different mechanism must therefore facilitate angular momentum transport in NGC 4593’s single-arm spiral.

Gravitational torques have been suggested to be a dominant source of torques in the intermediate scale (10\,pc-1\,kpc) region of AGN host galaxies \citep[e.g.][]{Hernquist:1989, Schlosman:1989, Shlosman:1990, Jogee:2006, Haan:2009, Hopkins&Quataert:2011}.
Despite high gas densities, the gravitational potential in NGC 4593's nuclear single-arm spiral is dominated by stars. 
Using a similar setup, \citet{Hopkins&Quataert:2010} used hydrodynamical simulations to show how on parsec scales, the $m=1$ mode arises from lopsided density distributions. It initially grows within the stellar component, which supports self-crossing orbits and is less influenced by the disc’s outer properties. In this configuration, stars act as an angular momentum sink, as interactions between gas streams and the lopsided stellar potential lead to angular momentum and energy loss \citep[e.g.][]{Chang:2007}, driving inward gas flow.
This mechanism propagates the $m=1$ mode from larger radii into the gravitational potential of the black hole. As shown in Sect.~\ref{subsec:Torques_Young_Stars}, young stars formed from recent gas inflow towards NGC 4593’s centre exhibit an off-centred distribution relative to the nucleus. This misalignment may reflect the non-axisymmetric gravitational potential characteristic of the $m=1$ mode. 

At $\sim$10–100\,pc from the centre lies the regime in which HQ10 proposed secondary instabilities linking kpc-scale host galaxy dynamics with (sub-)pc nuclear scales. Such instabilities can manifest as nuclear spirals, bars, rings, barred rings, or, as observed in \ngc, one-armed spirals. HQ10 describe how gravitational torques from lopsided, non-axisymmetric features can trigger and sustain $m=1$ gas inflows. Within this framework, the young stellar component acts both as a product of recent molecular gas inflow and as a sink for angular momentum from newly infalling gas. This mechanism only requires high central gas densities with lopsided distributions to generate the necessary torques.
However, it is important to note the scale differences between the HQ10 simulations and observations of \ngc. The kpc-scale gas disc is strongly self-gravitating. Only near the black hole SOI do the orbits become quasi-Keplerian, with the gravitational potential becoming spherical (dominated by the black hole), allowing $m=1$ features to stabilize as standing waves. While HQ10 suggest that $m=1$ features can extend beyond the black hole SOI, reaching radii of $\sim$50\,pc for a $3\times 10^7\,{\rm M}_\odot$ black hole, the single-arm spiral in \ngc is significantly larger.

\subsubsection{Triggering the gas instability}
\label{Sec:Complementary_processes}

Given that \ngc's nuclear single-arm spiral extends over kiloparsec scales, it is likely that, in addition to gravitational torques from the young stellar component, other mechanisms contribute to the transport of gas towards the centre. Angular momentum can be redistributed through interactions with the surrounding environment, such as perturbations in the gravitational potential or encounters with external perturbers.
The presence of a companion galaxy at a projected distance of $\sim$10\,kpc suggests that a past interaction could have triggered the initial gas instability. However, since the single-arm spiral in \ngc has a radius of only 1.9\,kpc -- much smaller than the distance to the companion -- such an interaction would require an extremely small impact parameter to be effective.

One of the main secular mechanisms to trigger gas inflow in barred galaxies, and fuel the very centre, is the bars within bars dynamical phenomenon \citep[e.g.][]{Schlosman:1989}. While gas might be stalled at the ILR of the primary bar, additional negative torques can then be produced by a nuclear bar located within the ILR. This has been shown to be effective, through nuclear spirals revealed in the molecular gas with ALMA \citep{Combes:2014, Audibert:2019, Audibert:2021}.
Nuclear spirals also emerge in the simulations by \cite{Emsellem:2015}, where the gas commonly forms one to three armlets. These structures arise from instabilities driven by interactions among gas, stars, and the central black hole, similar to those seen in simulations where bars and central masses propagating spirals to small scales \citep[e.g.][]{Englmaier&Shlosman:2000, Maciejewski:2004}. Stellar feedback from supernovae explosions and radiative pressure expels and re-accretes gas, redistributing angular momentum. These processes, along with gravitational torques and clump interactions (chaotic cold accretion, \citealt{Gaspari:2017, Wittor:2020}) ensure that the gas distribution remains highly dynamic, consistent with the non-stable, evolving nature of the single-arm spiral in \ngc.
Together, these processes -- gravitational torques, external interactions, and chaotic cold accretion -- likely play complementary roles in driving the sub-kiloparsec scale gas dynamics observed in \ngc’s nuclear single-arm spiral.

\subsection{Inflow mass rate and BHAR}
\label{subsec:Dsicussion_Mass_Transport_Rates}

To assess the importance of the $m=1$ mode for growing \ngc's SMBH, we estimate how the nuclear single-arm gas spiral can sustain the present-day AGN accretion rate.
The BHAR can be estimated from $M_{\rm BH}$ and $\lambda_{\rm Edd}$ assuming $L_{\rm Edd} = 1.5\times10^{38} (M_{\rm BH}/M_\odot)\, {\rm erg/s}$ for solar-composition gas and an accretion disc radiative efficiency of $\eta=0.1$, for the optically thick, geometrically thin accretion disc typical of luminous AGNs at these accretion rates \citep{Davis&Laor:2011}.
Adopting \ngc's bolometric AGN luminosity $L_{\rm bol} = 4.4 \times 10^{43}\,{\rm erg/s}$ estimated from its H$\beta$ luminosity \citep{Husemann:2022}, and $M_{\rm BH} = 4.47 \times 10^6\,{\rm M}_\odot$, the resulting Eddington ratio of $\lambda_{\rm Edd}=0.06 \pm 0.02$ translates to a BHAR of $0.10 \pm 0.03 \,{\rm M}_\odot/{\rm yr}$. 

From the host galaxy side, the net mass transport rates within \ngc's central 1.3\,kpc are negative across the entire single-arm spiral, with a radially averaged value of $-4.2 \,{\rm M}_\odot\,{\rm yr}^{-1}$ (see Sect~\ref{subsec:kinematic_modelling_mol_gas}).
We assume that only a small fraction of the inflowing gas fuels the AGN, $\sim$2\% of the gas mass, consistent with the present state. The majority is consumed by star formation \citep{Hopkins&Quataert:2010}, AGN- and star-formation-driven outflows, and cloud interactions that facilitate angular momentum removal. 
However, in \ngc, the currently observed SFR is much lower than the inflow rate, suggesting that a significant fraction of the gas is not immediately forming stars, possibly due to suppressed or delayed SF.
Consequently, the molecular gas inflow rate naturally exceeds the BHAR by approximately one and a half orders of magnitude.
With these assumptions, the nuclear single-arm spiral hosts sufficient gas to maintain the BHAR for 35\,Myr through the $m=1$ mode, resulting in a net growth of $3.5\times 10^5\,\mathrm{M}_\odot$, or 9\% of the current SMBH mass.
Uncertainties include a potential underestimation of the molecular gas mass in \ngc's nuclear single-arm spiral due to interferometric observations potentially missing larger-scale, extended emission. 
Additionally, the assumption of a relatively low value for the CO-to-H$_2$ conversion factor $\alpha_{\rm CO}$ could further underestimate the molecular gas masses (see Sect.~\ref{subsec:Data_ALMA}).
As $\alpha_{\rm CO}$ increases, also $\Sigma_{\rm mol}$ and its angular momentum grow linearly, while torques and mass inflow rates stay unchanged. This results in the gas reservoir's "lifetime” increasing linearly with $\alpha_{\rm CO}$. If, instead of the default conservative value (Sect.~\ref{subsec:Data_ALMA}), we adopt an optimistic value of $\alpha_{\rm CO} = 1/2 \alpha_{\rm CO}^{\rm MW}$, the nuclear single-arm spiral provides fuel to grow \ngc's \mbh by 45\% over 175\,Myr.

We note that this estimate relies on simplistic assumptions. While useful as an order-of-magnitude estimate, the BHAR can vary significantly on timescales of 10$^4$–10$^6$ years \citep{Shen:2007, Eftekharzadeh:2015, Khrykin:2021}, so the inflow rate measured today on parsec scales may not directly correspond to the nucleus's accretion rate. 
Furthermore, the comparison is based on a closed-box model, whereas \ngc's nuclear single-arm spiral does clearly not suffice the assumption of a closed system. 
While outflows may remove only a small amount of mass, the galaxy-scale bar likely provides a continuous gas supply to the outer end of the spiral. Indeed, \cite{Diaz-Garcia:2021} found that 1.6\,$\times$ more molecular gas is distributed along \ngc's galaxy-scale bar compared to the innermost 3.6\,kpc. In bar-dominated galaxies, gas is efficiently funnelled inward along bar dust lanes \citep{Athanassoula:1992, Kim:2012, Sormani:2015, Sormani:2023}, suggesting that the single-arm spiral could receive a sustained gas supply over much longer timescales than the estimated 35\,Myr "lifetime” of \ngc's nuclear single-arm spiral.

\subsection{Possible implications for cosmic black hole growth} 
\label{subsec:Discussion_Fuelling_Mechanism}

As discussed in Sect.~\ref{Sec:Complementary_processes}, the single-arm spiral may not be stable or as pronounced in its present-day configuration 
Nevertheless, the $m=1$ mode appears to play a critical role for \ngc's AGN as a self-sustained mechanism enabling continuous and steady gas accretion rates over megayear timescales. Furthermore, it can rely on a persistent gas supply from scales beyond 1\,kpc -- channelled via the bar dust lanes -- to sustain this process over even longer periods. This makes the $m=1$ mode a potential missing link on intermediate scales, bridging the kpc-scale host galaxy dynamics with the black hole-dominated spherical gravitational potential in galactic centres.

In the context of the overall AGN population, there is no compelling reason to consider the processes in \ngc unique. While the outstanding dust absorption and CO emission are a textbook example of the expected observational signatures of the $m=1$ mechanism, the galaxy was selected purely based on its \mbh, specific accretion rate, and high central gas densities.
If similar features are common in rigorously selected luminous type-1 AGNs, the $m=1$ mode could play a role in the build-up of cosmic SMBH mass in the highest-accreting AGNs.
This would complement the range of processes observed in nearby low-$L$ AGNs, whose accretion rates are generally too low to explain SMBH growth on relevant timescales.
In luminous AGNs -- responsible for the majority of black hole mass growth since $z=2$ -- fueling is not driven by major galaxy mergers, which are both rare  \citep{Lotz:2011} and short-lived \citep{Cisternas:2011, Kocevski:2012, Mechtley:2016}. Instead, half of the AGNs responsible for black hole growth since $z=1$ reside in disc-dominated galaxies, indicating a lack of recent strong interactions. This suggests that secular processes or internal instabilities are the primary mechanisms channelling gas from kiloparsec to sub-parsec scales.
In this context, the $m=1$ mode, as observed in NGC 4593, may contribute meaningfully to SMBH growth in the Universe.

A challenge to this interpretation is that similarly prominent sub-kiloparsec $m=1$ patterns are rarely observed in the centres of nearby galaxies \citep{Phookun:1993, Emsellem:2001, Schinnerer:2002}. More commonly, central regions exhibit resonances where gas accumulates and triggers star formation \citep{Mazzuca:2008, Comeron:2014}, often associated with two-armed spirals ($m=2$; \citealt{Englmaier&Shlosman:2000, Maciejewski:2002, Ann&Thakur:2005, Combes:2014, Liang:2024}) or more complex patterns \citep{Gadotti:2019, Schinnerer:2023}.
However, in these nearby galaxies the BHAR, typically inferred from lower-luminosity type-1 AGNs or uncertain diagnostics in type~2 AGNs, tend to be low ($\lambda_{\rm Edd} << 0.1$). The relative scarcity of luminous type-1 AGNs, together with their greater distances, limits spatial resolution, making it difficult to measure sub-kiloparsec gas dynamics.
This observational bias might explain why clear signatures of central $m=1$ mode, whether in dust absorption or gas emission, have not been more frequently observed in the local AGN population.
Future high-resolution observations of luminous AGNs will be crucial for assessing how widespread and significant this mechanism is in shaping SMBH evolution across cosmic time.

\section{Summary and conclusions}
\label{sec:Summary&Conclusions}

In this work, we have presented the results of a systematic search for a secular AGN feeding mechanism. The target galaxy was selected for its high central gas densities and accretion rates, representative of cosmic black hole growth over the past 10 billion years of the Universe. 
We identified \ngc\ as a case study and investigated the intrinsic mechanisms driving angular momentum transport and gas inflow in its nuclear region. Through an analysis of molecular gas dynamics and extended UV emission, we identified a key mechanism regulating gas inflow from kiloparsec scales down to the black hole SOI.
Our primary results are summarised as follows:

\begin{itemize}

    \item Molecular gas with a total mass of ${\rm log}\, M_{\rm mol} / {\rm M}_\odot = 8.1 \pm 0.3$ is concentrated in a prominent single-arm spiral that is also visible in dust absorption.

    \item Despite modest star formation along the spiral (integrated ${\rm SFR}=4.9 \times 10^{-2} \, \mathrm{M}_\odot/\mathrm{yr}$), the low star formation efficiency suggests that most of the gas is funnelled towards feeding the AGN rather than forming stars.
    
    \item The inflowing gas has built up a young stellar component in the centre of \ngc, offset from the AGN by $258 \pm 14 \,{\rm pc}$ and with a mass of ${\rm log} M_\star / {\rm M}_\odot = 7.5 - 9.3$. The resulting lopsided gravitational potential generates torques that alone are sufficient to drive the kinematically inferred molecular gas inflow, although the current configuration represents only a snapshot in time.

    \item On several $\times 10$\,pc scales, our observations qualitatively support the gas transport mechanism proposed by \cite{Hopkins&Quataert:2010}, in which gravitational torques from lopsided central gas distributions trigger instabilities and form $m = 1$ features. However, in \ngc, the nuclear single-arm spiral extends well beyond the SOI of the black hole, where the configuration is likely more dynamic and variable.
    
    \item The molecular gas mass reservoir and mass inflow rates along the single-arm spiral can sustain AGN fuelling for at least 35\,Myr, enabling $\sim$10\% growth of \ngc's central SMBH at the current accretion rate. With gas from the dust lanes continuously feeding the outer end of the nuclear spiral, this process could persist even longer.
    
\end{itemize}

In summary, our study of \ngc demonstrates how the $m=1$ mode, in the form of a single-arm gas-rich spiral, transports gas from galactic scales down to the black hole SOI. 
This indicates that the $m=1$ mode may represent a fundamental galaxy-intrinsic AGN feeding mechanism operating on intermediate scales -- connecting gas flows from kiloparsec-scale host galaxy structures down to black hole-dominated scales. With gas inflow rates sufficient to grow the central SMBH over the AGN lifetime, this process may play a significant role in the growth of \ngc's central black hole.
To better understand the relevance of this mechanism, it is essential to investigate whether the $m=1$ mode is a common or rare phenomenon among luminous AGNs accreting at similar rates. A census of nuclear single-arm structures -- through dust absorption, ionised gas emission, or molecular gas emission in AGN-hosting galaxies -- will help determine the contribution of this mechanism for growing the overall SMBH population since cosmic noon.

\begin{acknowledgements}
We would like to warmly thank Eric Emsellem for useful input which helped the writing of this paper.
TAD acknowledges support from the UK Science and Technology Facilities Council through grants ST/S00033X/1 and ST/W000830/1.
VNB gratefully acknowledges support through the European Southern Observatory (ESO) Scientific Visitor Program. 
MG acknowledges support from the ERC Consolidator Grant \textit{BlackHoleWeather} (101086804).
JAFO acknowledges financial support by the Spanish Ministry of Science and Innovation (MCIN/AEI/10.13039/501100011033), by ``ERDF A way of making Europe'' and by ``European Union NextGenerationEU/PRTR'' through the grants PID2021-124918NB-C44 and CNS2023-145339; MCIN and the European Union -- NextGenerationEU through the Recovery and Resilience Facility project ICTS-MRR-2021-03-CEFCA.
This paper makes use of the following ALMA data: ADS/JAO.ALMA\#2017.1.00236.S and ADS/JAO.ALMA\#2018.1.00978.S. ALMA is a partnership of ESO (representing its member states), NSF (USA) and NINS (Japan), together with NRC (Canada), MOST and ASIAA (Taiwan), and KASI (Republic of Korea), in cooperation with the Republic of Chile. The Joint ALMA Observatory is operated by ESO, AUI/NRAO and NAOJ.\\
The National Radio Astronomy Observatory is a facility of the National Science Foundation operated under cooperative agreement by Associated Universities, Inc.
This work is based on observations with the NASA/ESA Hubble Space Telescope obtained from the Data Archive at the Space Telescope Science Institute, which is operated by the Association of Universities for Research in Astronomy, Incorporated, under NASA contract NAS5-26555. Support for Program number HST-AR 17063 (PI Bennert) was provided through a grant from the STScI under NASA contract NAS5-26555.\\
Based on observations collected at the European Organisation for Astronomical Research in the Southern Hemisphere under ESO programmes 099.B-0242(B) and 0103.B-0908(A).
\end{acknowledgements}

%
%

\bibliographystyle{aa}
\bibliography{references}

\begin{appendix}

\end{appendix}

\end{document}